\documentclass[a4paper,11pt]{article}
\pdfoutput=1 
\usepackage{jheppub} 
\usepackage[T1]{fontenc} 
\usepackage[page]{appendix}
\usepackage{amsmath}
\usepackage{amssymb}
\usepackage{amsfonts}
\usepackage{xcolor}

\newcommand{\sg}[1]{\mathsf{#1}}

\usepackage{subfig}
\usepackage{graphicx}
\usepackage{breqn}
\usepackage{bm}
\usepackage[left]{lineno}
\usepackage{yfonts}

\title{\boldmath New description of neutrino flavour evolution  in solar matter}

\author[a]{J. Rembieli\'{n}ski}
\author[b,1]{J. Ciborowski,\note{Corresponding author.}}

\affiliation[a]{University of {\L}\'od\'z, Faculty of Physics and Applied Informatics, Pomorska 149/153, PL-90236 {\L}\'od\'z, Poland}
\affiliation[b]{University of Warsaw, Faculty of Physics, Pasteura 5, PL-02-093 Warsaw, Poland}

\emailAdd{jaremb@uni.lodz.pl}
\emailAdd{cib@fuw.edu.pl}

\abstract{Assuming that  the interacting neutrino and  the solar matter  can be treated   as an open quantum system
we formulate a formalism of  neutrino state evolution  in terms of the Bloch vector,  obtained by solving  a quasilinear extension of the  von Neumann  equation.
We broaden  the   classical  Wolfenstein formalism by means of  exploiting   the  strictly  linear von Neumann equation  and  reinterpreting the results in terms of the quantities appropriate for the quasilinear approach. We obtain  similar predictions for  the averaged neutrino survival probability measured on Earth in  both approaches,  differing  by details inside  the Sun where the quasilinear evolution  predicts a suppression  of oscillations. We discuss the issues of   energy transfer between the neutrino and the Sun and the speed of the state evolution in both approaches.}

\begin{document}
\maketitle
\flushbottom


\section{Introduction}\label{sec:Introduction}

The fact that the  pattern of  neutrino oscillations in matter   deviates from that in the vacuum  was recognised and mathematically described  by Wolfenstein more than  40 years ago. His formalism  was  based on an assumption  of a  linear  evolution of the  neutrino state, involving a free Hamiltonian and the so-called  Wolfenstein potential (proportional to the electron number density)  as  the interaction  term~\cite{cite:Wolfenstein78,cite:Wolfenstein79}.
In the 80-ties this topic was expanded  by Mikheev and Smirnov who also reestablished  a  new  nomenclature, such as   "resonance enhancement of oscillations" and "adiabatic flavour conversion",  as well as   emphasised the role of effective flavour  mixing (mentioned already by Wolfenstein),  induced by the varying Wolfenstein potential  in the course of neutrino evolution in the solar matter~\cite{cite:MS1,cite:MS2,cite:MS3,cite:MS4,cite:MSW1}. The MSW effect, linked to this formalism,  explained  the deficit of $\nu_e$'s in the solar flux measured on Earth, compared to predictions of the Standard Solar Model,  and also allowed to calculate  corrections to neutrino oscillations in the terrestrial matter.
We now note that  one circumstance  should be  taken into account in  developing an  appropriate evolution formalism.   Namely, in reality the propagating neutrino interacts with the solar plasma, rather not with a potential field, in a way understood for an open system, i.e., subsystem (neutrino) and environment (plasma), respectively. Thus we can expect that a  complete description of the neutrino flavour state can be obtained by solving an appropriate master evolution equation describing open systems.

The main aim of the present  paper is to  formulate  an  approach to neutrino flavour evolution in the solar matter in terms of  a  quasilinear extension of the Giorini– Kossakowski–  Sudarshan– Lindblad (GKLS) master equation pertaining to  open systems~\cite{cite:GKLS1} -- one of the most general types of equations describing  evolution of the density matrix   that preserves the laws of quantum mechanics. We  use it in its simplest  form,  reduced to  a quasilinear von Neumann equation~(\ref{eq:vonNeumann2}),  of which the linear von Neumann equation~(\ref{eq:vonNeumann1}) is a special case. We thus  distinguish  a strictly linear (SL)  and  a  quasilinear (QL)  description of neutrino flavour  evolution  within this general framework.

The  principal results of our research are as follows.
Firstly,   the dynamics  of the state evolution  in both discussed cases  are determined by one variable  related to the instability point of the evolution equation. We have thus obtained a  neat  interpretation  of the quantities describing  both  oscillatory and non-oscillatory neutrino state evolution in the solar matter.
Secondly,  both  evolutions lead to the same predictions as regards  the survival probability  of neutrinos from  the higher end of the solar energy spectrum measured  on Earth, however  the forms of evolution   contrast  substantially  in   the solar core where oscillations are totally suppressed according to the quasilinear description. Thirdly, we supplement both approaches  with the results for the neutrino-Sun energy transfer as well as the speed of the neutrino state evolution.

\section{Theoretical setting}\label{sec:Theor}

The role  of  matter  in   neutrino state evolution stems from the  uniqueness  of the CC interactions of the electron neutrinos.
The exceptionally large magnitude of this effect in the solar matter is due  to the fact that the Sun  consists  of a high density
plasma containing electrons.
In the following we elaborate on two contrasting   descriptions of neutrino state evolution  in this environment, both within the density matrix formalism involving the Bloch vector, denoted  $\bm{n} = (n_1,n_2,n_3)$.  The  first is  based on the strictly  linear von Neumann  evolution equation~(\ref{eq:vonNeumann1})  with  the interaction term (potential) contained in the Wolfenstein Hamiltonian~(\ref{eq:WolfensteinH}). The second one is based on  a natural, nonlinear but quasilinear extension of the von Neumann evolution equation~(\ref{eq:vonNeumann2}) with the free Hamiltonian and interactions incorporated  in an additional   evolution generator governing the environment reaction to the neutrino propagation. This is a novel, unified approach in which  the MSW  'resonance' point  is related to   a structural instability point characteristic of the quasilinear  evolution equation. We  consider   neutrinos  produced in the centre of the Sun  with  the subsequent  radial motion    parameterised by the  travelled  distance, $L$, instead of time,  owing to  the small neutrino masses.

\subsection{Linear evolution}\label{sec:Theor-linear}

Wolfenstein  proposed  to account for flavour-originating difference of neutrino interactions with matter  by means of  a position-dependent potential, $V_e(L)=\sqrt{2}{\rm G_F} n_e(L)$, where  $n_e(L)$  is the  electron number density at a distance $L$ from the centre of the Sun and  ${\rm G_F}$   is  the Fermi constant~\cite{cite:Wolfenstein78,cite:Wolfenstein79}.
In consequence, the Wolfenstein's mathematical description of the strictly linear evolution  was based on   the Schroedinger-like equation
with the  Hamiltonian  modified for the  presence of matter  by means of   the above  potential term. The full  Hamiltonian, $H_{\rm tot}$,
in the flavour basis for  two-flavour  oscillations   reads (cf. Appendix~I)
\begin{linenomath}
\begin{equation}\label{eq:WolfensteinH0}
H_{\rm tot}(L) = \Bigg (E +\frac{V_e(L)}{2} \Bigg )\cdot I + H_W(L)
\end{equation}
\end{linenomath}
with
\begin{linenomath}
\begin{gather}\label{eq:WolfensteinH}
H_W(L) = \frac{1}{2}
\begin{pmatrix}
-\tfrac{\Delta m^2 }{2E}\cos 2\theta + V_e(L) \;\;\;\;\;& \tfrac{\Delta m^2 }{2E}\sin 2\theta\\
\tfrac{\Delta m^2}{2E}\sin 2\theta \;\;\;\;\;& \tfrac{\Delta m^2 }{2E}\cos 2\theta - V_e(L)
\end{pmatrix},
\end{gather}
\end{linenomath}

where  $\Delta m^2 =\Delta m_{21}^2 $
and $E$ is the arithmetic mean of the neutrino kinetic  energies,  $E_i=\sqrt{p^2+m_i^2}$, in the mass  basis spanned by the free neutrino energy eigenstates $\nu_i, i=1,2$. We assume the  neutrino momentum squared, $\bm{p}^2=p^2$, to be   fixed during evolution  while  $\Delta m^2=m_2^2 - m_1^2 = E_2^2 - E_1^2 = 2 E \Delta \!E$, where  $\Delta \! E= E_2 - E_1$  (cf. Appendix~I)
and $\theta$ is the mixing angle  ($\theta=\theta_{12}$) in the vacuum.
Eigenvalues of the  Hamiltonian~(\ref{eq:WolfensteinH}), $\lambda_{\pm}$,  are given by
\begin{linenomath}\begin{equation}\label{eq:EigenValuesH}
\lambda_{\pm} (L) = \pm \sqrt{ \left (\frac{\Delta m^2}{2E}\sin 2\theta \right )^2 + \left (V_e(L)-\frac{\Delta m^2}{2E}\cos 2\theta \right )^2}.
\end{equation}\end{linenomath}

Now, in distinction  to  most of the earlier treatments of  the subject, we develop a  description of neutrino oscillations in the solar matter according to the formalism  of the density matrix reduced to the flavour space, $\rho (L)$,  which provides a complete description of the  flavour state along the neutrino trajectory in the Sun.
The density matrix  is  chosen in the form
\begin{linenomath}\begin{equation}\label{eq:RhoNeutrinos}
\rho(L) = \frac{1}{2} \Big(I+\bm{n}(L)\bm{\sigma} \Big),
\end{equation}\end{linenomath}
where  $\bm{\sigma}$ denotes the Pauli matrices and $\bm{n}$ ($\bm{n}^2\leq 1$) is the Bloch vector.
The evolution of $\rho (L)$ is governed by the strictly linear  von Neumann  equation  (hereafter we use $\hbar=c=1$)
\begin{linenomath}\begin{equation}\label{eq:vonNeumann1}
\rho^{\,\prime}(L) =-i [H_{\rm W} (L),\rho (L)],
\end{equation}\end{linenomath}
with an initial condition $\rho(0)=\rho_0$, where only  the oscillation part, $H_{\rm W}(L)$,   of the full Hamiltonian~(\ref{eq:WolfensteinH0}) participates since the term proportional to the identity matrix  drops out in the commutator;  the prime ($\prime$) denotes differentiation over $L$.
For pure neutrino states which we consider here,  Eq.~\ref{eq:vonNeumann1}  is equivalent to the well-known Wolfenstein equation~\cite{cite:Wolfenstein78,cite:Wolfenstein79}.
In terms of the Bloch vector it  reduces to  the form
\begin{linenomath}\begin{equation}\label{eq:BlochEquationSL}
\bm{n}^{\prime}(L) = \bm{\omega}(L) \times \bm{n}(L)
\end{equation}\end{linenomath}
with  the flavour  vector $\bm{\omega}$,  according to~(\ref{eq:WolfensteinH}), given by
\begin{linenomath}\begin{equation}\label{eq:VectorOmega1}
\bm{\omega}(L) =\bm{\omega}_{\rm W}(L) = \left (\frac{\Delta m^2 }{2E}\sin 2\theta,\:0,\:V_e(L)-\frac{\Delta m^2}{2E}\cos 2\theta \right).
\end{equation}\end{linenomath}
The initial condition in~(\ref{eq:BlochEquationSL}) corresponds to  the electron neutrino  created  in the centre of the  Sun, $\bm{n}(0)=\bm{\nu}_e$, where $\bm{\nu}_e=(0,0,1)$  has a unit norm. Since Eq.~\ref{eq:BlochEquationSL} is norm-preserving,   $\bm{n}(L)^2=1$  during the entire  evolution
(for completeness,  the state   $\bm{\nu}_{\mu}=(0,0,-1)$  corresponds to the muon neutrino).
We adopt  a similar notation  corresponding to  mass eigenstates $\nu_1$ and  $\nu_2$,    $\bm{\nu}_1=(-\sin 2\theta,0,\cos 2\theta)$  and $\bm{\nu}_2=(\sin 2\theta,0,-\cos 2\theta)$, respectively.
Note that Eq.~\ref{eq:BlochEquationSL} is covariant under rotations of the vectors $\bm{\omega}$ and $\bm{n}$,  being the adjoint action of the  flavour group $\sg{SU(2)}$.
In the following  we  proceed with the above  considerations  into   a more general framework of a quasilinear  evolution, introduced recently elsewhere  by one of us~\cite{cite:RC1}.

\subsection{Quasilinear evolution}\label{sec:Theor-quasilinear}

Quantum operations  satisfying the so called quasilinearity  property, although can be  in general nonlinear, preserve the structure of  quantum ensembles and thus are admitted by the quantum mechanical laws (as well as  guarantee infeasibility of superluminal communication~\cite{cite:RC2}).  The  starting point for the following considerations is a  particularly simple,  quasilinear  extension of the GKLS equation~\cite{cite:AL1}, which involves  terms representing  the  interaction between  the subsystem and the environment and  has  the following  form~\cite{cite:RC1}
\begin{linenomath}\begin{equation}\label{eq:vonNeumann2}
\rho^{\,\prime} = -i [H,\rho] + \{G,\rho\} -2 \rho {\rm Tr}(G\rho),
\end{equation}\end{linenomath}
where $G$ is, in general $L$-dependent,  Hermitian matrix, in  the units  of energy like  the Hamiltonian. Eq.~\ref{eq:vonNeumann2} is a nonlinear but quasilinear extension of the von Neumann equation~(\ref{eq:vonNeumann1}), considered  in the present framework for the first time (it  has already  been  discussed in various other contexts ~\cite{cite:RC1,cite:RC2,cite:KR1,cite:Gisin1,cite:Brody1,cite:Kawabata1}, sometimes as an equivalent nonlinear Schroedinger equation for pure states).
The second term on the right-hand side of  Eq.~\ref{eq:vonNeumann2} (the so called "dissipator")  is responsible for gain and/or loss reaction of the environement~\cite{cite:Brody1}  while the last term provides the overall conservation of probability. Indeed,  the quasilinear evolution according to~(\ref{eq:vonNeumann2})  is trace-preserving so it does not violate the  probabilistic interpretation of the resulting description.
Moreover, one can also show, using ${\rm Tr}\,\rho=1$, that~(\ref{eq:vonNeumann2}) is invariant under the replacement $H\rightarrow H+f_1I$ and $G\rightarrow G+f_2 I$, where $I$ is the unit matrix and $f_1, f_2$ are arbitrary differentiable functions of $L$.
In the two-flavour approximation  we choose $H$ and $G$ traceless,  i.e.,
\begin{linenomath}\begin{equation}\label{eq:ForNeutrinos}
H=\frac{1}{2} \bm{ \omega\sigma} \;\;\;\;\;\;\;\; G=\frac{1}{2} \bm{ g \sigma},
\end{equation}\end{linenomath}
where $\bm{\omega}$,  $\bm{g}$ are real, in  general $L$-dependent, flavour vectors.
Eq.~\ref{eq:vonNeumann2} is form-invariant under the $\sg{SL(2,C)}$ group transformation defined as
\begin{linenomath}\begin{equation}\label{eq:SL}
\tilde{H} + i \tilde{G} = A (H + i G) A^{-1}
\end{equation}\end{linenomath}
together with
\begin{linenomath}\begin{equation}\label{eq:vonNeumannSolution0}
\tilde{\rho}  = \frac{A\rho A^{\dag}}{{\rm Tr}(A\rho A^{\dag})},
\end{equation}\end{linenomath}
where $A\in \sg{SL(2,C)}$.
Now, by means of~(\ref{eq:RhoNeutrinos}), (\ref{eq:vonNeumann2}) and  (\ref{eq:ForNeutrinos}) one obtains the corresponding equation for the Bloch vector
\begin{equation}\label{eq:BlochEquation}
\bm{n}^{\,\prime}(L) = \bm{\omega} \times \bm{n}(L) +  \bm{g}  -  \Big(  \,\bm{g} \bm{n}(L)\,\Big) \bm{n}(L),
\end{equation}
with the same  initial condition as in the linear case,  $\bm{n}(0)=\bm{\nu}_e$.  For $\bm{g}=0$ and $\bm{\omega}=\bm{\omega}_{\rm W}$   one recovers the linear equation~(\ref{eq:BlochEquationSL}).  Eq.~\ref{eq:BlochEquation} is autonomous and  belongs to the class of (inhomogeneous) Riccati systems  -- first order ordinary differential equations,  quadratic in the unknown vector function $\bm{n}(L)$.
It preserves  purity of the quantum state, i.e., the condition $\bm{n}(L)^2=1$ is  invariant under this evolution.
Moreover, it has been shown~\cite{cite:RC1,cite:KR1} that  such a  system has points of structural instability~\cite{cite:Arnold1,cite:Wiggins1}, not related to initial conditions. This means  that small changes  of the system parameters near these  points can lead to significant modifications in the overall dynamics.
In order to gain some insight into the quasilinear evolution, one can  solve~(\ref{eq:vonNeumann2}) analytically  in a simple case of
distance-independent generators $H$ and $G$, as was shown elsewhere~\cite{cite:RC1,cite:KR1}.  To a good approximation, this would also be the case for  oscillations in   terrestrial matter  with  a constant density ($V_e\approx {\rm const}$).
One yields in this case
\begin{linenomath}\begin{equation}\label{eq:vonNeumannSolution}
\rho (L)  = \frac{K(L)\rho (0) K^{\dag}(L)}{{\rm Tr}\Big(K(L)\rho (0) K^{\dag}(L)\Big)},
\end{equation}\end{linenomath}
where
$K(L)=e^{L(G-i H) }$.
Thus if the  generator $H+iG$ is nilpotent, the solution~(\ref{eq:vonNeumannSolution}) differs radically from the case  when the nilpotency condition does not hold  so this feature  fixes the structural instability of this evolution. Consequently, for more general  $L$–dependent generators  $H$  or  $G$  one   should expect that if  $H(L)+i G(L)$ is nilpotent  for a specific  $L=L_{\rm ins}$, this  point on the trajectory fixes a structurally unstable point of evolution of the density matrix, $\rho(L)$.  In the two-flavour case one has
\begin{linenomath}\begin{equation}\label{eq:HiG}
H(L) + i G(L) = \tfrac{1}{2} \Big (  \bm{\omega}(L) + i \bm{g}(L)  \Big ) \bm{\sigma}
\end{equation}\end{linenomath}
so the nilpotency of~(\ref{eq:HiG}) implies that a structural instability arises for $L=L_{\rm ins}$, where  the vectors $\bm{\omega}(L_{\rm ins})$ and $\bm{g}(L_{\rm ins})$ have equal lengths and are perpendicular, i.e.,
\begin{linenomath}\begin{equation}\label{eq:C1C2a}
\bm{\omega}(L_{\rm ins})^2 = \bm{g}(L_{\rm ins})^2 \;\;\;\;\;\;\;\;  \bm{\omega}(L_{\rm ins}) \bm{g }(L_{\rm ins})=0.
\end{equation}\end{linenomath}
Since
\begin{linenomath}\begin{equation}\label{eq:C1C2}
C_1 = \bm{\omega}^2 - \bm{g}^2  \;\;\;\;\; C_2 = \bm{\omega g}
\end{equation}\end{linenomath}
are Casimir invariants of the $\sg{SL(2,C)}$ transformations~(\ref{eq:SL}),
conditions~(\ref{eq:C1C2a}) are $\sg{SL(2,C)}$-invariant and correspond to $C_1(L_{\rm ins}) = C_2(L_{\rm ins})=0$.
Consequently, the generator $\frac{1}{2}  (\bm{\omega}+i\bm{g})\bm{\sigma}$ is  nilpotent  for $C_1=C_2=0$.
Thus in a particular  case of constant $\bm{\omega}$ and $\bm{g}$  the corresponding evolution is necessarily rational
while for other values of $C_1$ and $ C_2$ it can be oscillatory, stationary or  have a mixed character~\cite{cite:RC1,cite:KR1}.

In order to quantify the influence of the structural  instability point on the general evolution of the density matrix  we introduce a corresponding measure  in the space of invariants $(C_1, C_2)$. This quantity, denoted by  $d(L)$,  will be termed "energy gap". It  has the dimension of energy and its square is defined as  the   square of the  distance   between configurations given by the pairs $\Big(C_1(L), C_2(L)\Big)$ and $\Big(C_1(L_{\rm ins}), C_2(L_{\rm ins})\Big)=(0,0)$.  Thus the energy gap reads
\begin{linenomath}\begin{equation}\label{eq:EnergyGap1}
d(L) = \sqrt{|C_1(L) + 2 i C_2(L)|} = \Big (C_1(L)^2 + 4 C_2(L)^2\Big )^{\frac{1}{4}}.
\end{equation}\end{linenomath}
The energy gap  will be shown to play a fundamental  role in describing  the evolution  of  the neutrino state in the solar matter.
In the considered case of   two-flavour oscillations with traceless generators $H$ and $G$, the energy gap  can be expressed by the eigenvalues, $\lambda (L)$, of the generator $H(L) + i G(L)$. Indeed, since it follows from (\ref{eq:HiG}), (\ref{eq:C1C2}) (cf. Appendix~I) that
\begin{linenomath}\begin{equation}\label{eq:eigenvalues1}
 \lambda(L) =  \pm \frac{1}{2} \sqrt{C_1(L) + 2 i  C_2(L)},
\end{equation}\end{linenomath}
the  energy gap is  given simply by
\begin{linenomath}\begin{equation}\label{eq:EnergyGap21}
 d(L) = 2 |\lambda(L)|
\end{equation}\end{linenomath}
(the eigenvalues $\lambda (L)$ vanish in the structural instability point, $\lambda(L_{\rm ins})=0$).
The formalism  described in  this subsection  is applicable   to the linear   case too, under the choice   $C_1=\bm{\omega}_{\rm W}^2$~(\ref{eq:VectorOmega1}) and  $C_2=0$~($\bm{g}=0$).

\section{Evolution of the neutrino state in the solar matter}\label{sec:evolution}

\subsection{Preliminaries}\label{subsection:Preliminaries}

The most precise  measurement of  $\Delta m_{21}^2=(7.54^{+0.19}_{-0.18}) \cdot 10^{-5}$ eV$^2$  was delivered by KamLAND~\cite{cite:KamLAND1}.
The past $2\sigma$ discrepancy between   the solar and KamLAND neutrino results has recently been significantly reduced owing to the inclusion of   new SK4 data  which yielded  $\Delta m_{21}^2=(6.11^{+1.21}_{-0.68}) \cdot 10^{-5}$ eV$^2$  for the SK+SNO datasets~\cite{cite:Nakajima1}.
In  numerical calculations we   use  the newest (2021) best fit  central values from  the global analysis of the neutrino data: $\Delta m^2=\Delta m_{21}^2=(7.42^{+0.21}_{-0.20} ) \cdot 10^{-5}$ eV$^2$ (dominated by the KamLAND measurement) and $\theta=\theta_{12}=(33.44^{+0.77}_{-0.74})^{\circ}$~\cite{cite:NuFit}~(www.nu-fit.org).
We  note that  the threshold energy  as well as the characteristic distances (see below) depend  on   the   value of $\Delta m^2_{21}$ and $\theta$.

The radial distribution of the electron number density, $n_e(L)$,   was adopted according to the  Standard  Solar Model  prediction for the solar matter density (BS2005-AGS,OP)~\cite{cite:SSMData}. The Wolfenstein potential, $V_e(L)=\sqrt{2}{\rm G_F} n_e(L)$,  where $\rm G_F$ is the Fermi constant, was  assumed in the form
$V_e(L) =   V_e(0)\, h_e(L)$  with $V_e(0) =  0.012$~neV and  the   following  effective parametrisation of the radial  dependence
\begin{linenomath}\begin{equation}\label{eq:InvMorse1}
 h_e(L)= 1 - \Big (1 - \exp(- L/L_0 ) \Big )^{\alpha}
\end{equation}\end{linenomath}
where  $L_0=64850$~km and $\alpha=2.1$~(cf. Fig.~\ref{fig:Fig1402l}).

The electron  neutrino survival probability, $p_{ee}$, is   expressed by the  third component of the Bloch vector,  $p_{ee}(L) = \frac{1}{2}\Big( 1+n_3 (L)  \Big)$  in  two-flavour approximation. Consequently, muon neutrino appearance probability is given by $p_{e\mu}(L) = \frac{1}{2}\Big( 1-n_3 (L)  \Big )$.

\subsection{Linear evolution of solar neutrinos}

For  the evolution of the Bloch vector governed by the strictly linear  equation~(\ref{eq:BlochEquationSL})  one obtains,  with the aid of~(\ref{eq:VectorOmega1}), the  following admissible configurations, $(C_1,C_2)$,  in the space of invariants
\begin{linenomath}\begin{equation}\label{eq:C1C2MSW}
C_{\rm 1,W}(L) = \bm{\omega}^2_{\rm W} = \left (\frac{\Delta m^2}{2E}\sin 2\theta \right )^2 + \left (V_e(L)-\frac{\Delta m^2}{2E}\cos 2\theta \right )^2
\;\;\;\;\;  C_{\rm 2,W}(L)=0
\end{equation}\end{linenomath}
and thus the  corresponding  energy gap~(\ref{eq:EnergyGap1})   reads
\begin{linenomath}\begin{equation}\label{eq:EnergyGapL}
d_{\rm W}(L) = \sqrt{ \left (\frac{\Delta m^2}{2E}\sin 2\theta \right )^2 + \left (V_e(L)-\frac{\Delta m^2}{2E}\cos 2\theta \right )^2}.
\end{equation}\end{linenomath}
Denoting by $L_{\rm res}$ the distance at which the energy gap  $d_{\rm W}(L)$  reaches  a  minimum, $d_{\rm W,min}$,
allows to write the corresponding relationship
\begin{linenomath}\begin{equation}\label{eq:ResCondition1}
V_e(L_{\rm res})=\frac{\Delta m^2}{2E}\cos 2\theta,
\end{equation}\end{linenomath}
which coincides  with the  'resonance' condition introduced  by Mikheev and Smirnov~\cite{cite:MS1,cite:MS2,cite:MS3,cite:MS4}.
The energy gap at the 'resonance' distance equals
\begin{linenomath}\begin{equation}\label{eq:C1C2GapMin}
 d_{\rm W}(L_{\rm res})= d_{\rm W,min} = \frac{\Delta m^2}{2E}\sin 2\theta.
\end{equation}\end{linenomath}
Thus   the structural  instability point is  inaccessible in the  case of  the  strictly linear  evolution~(\ref{eq:BlochEquationSL}), since $C_{\rm 1,W}$ cannot equal
zero for $\theta\ne 0$~(\ref{eq:C1C2MSW});  nevertheless one can expect an influence of this point on the evolution dynamics.
Since the  electron number density in the Sun  monotonically decreases to zero with increasing $L$, Eq.~\ref{eq:ResCondition1}  can be numerically solved for  $L_{\rm res}$.
The neutrino energy must exceed the threshold value, $E_{\rm th, W}$,  for the  'resonance'  point to exist, determined by the condition $\Delta m^2/2E < V(0) $, i.e.,
\begin{linenomath}\begin{equation}\label{eq:ThresholdEnergy1}
E_{\rm th,W} =  \frac{\Delta m^2}{2 V_e(0)}\cos 2\theta \approx 1.21~{\rm MeV}.
\end{equation}\end{linenomath}
One obtains $L_{\rm res}\approx 182800$~km for $E=10$~MeV and
$L_{\rm res}\approx 135500$~km for $E=5$~MeV.
The value of the energy gap  at the  'resonance'   distance, $d_{\rm W,min}$,  amounts to 0.0034~neV and 0.0068~neV, respectively,
therefore the influence of the 'resonance' point  at a higher energy is stronger than at a lower.

\begin{figure}[!htbp]
    \begin{center}
        \includegraphics[width=0.45\linewidth]{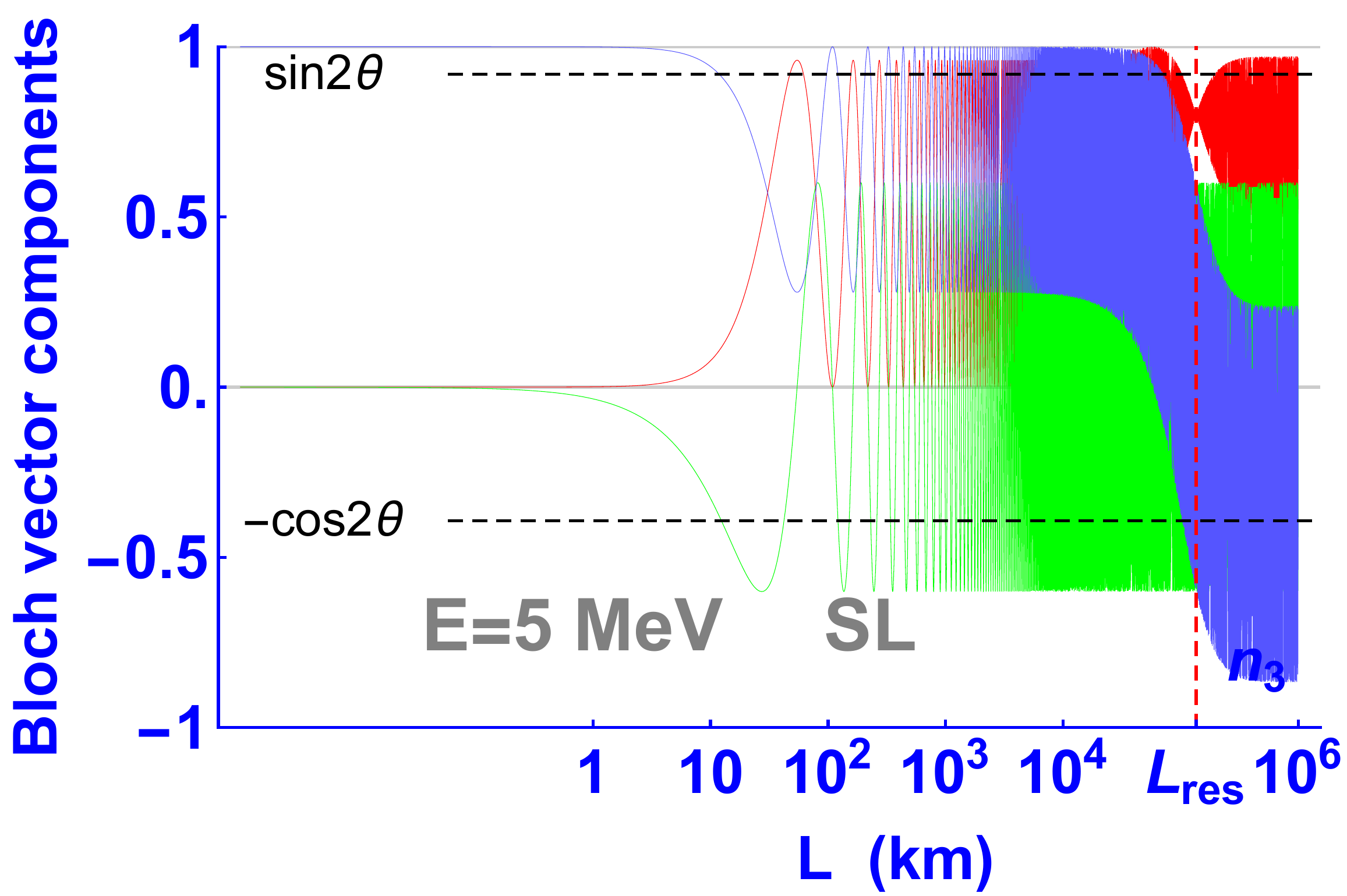}
        \hfill
        \includegraphics[width=0.45\linewidth]{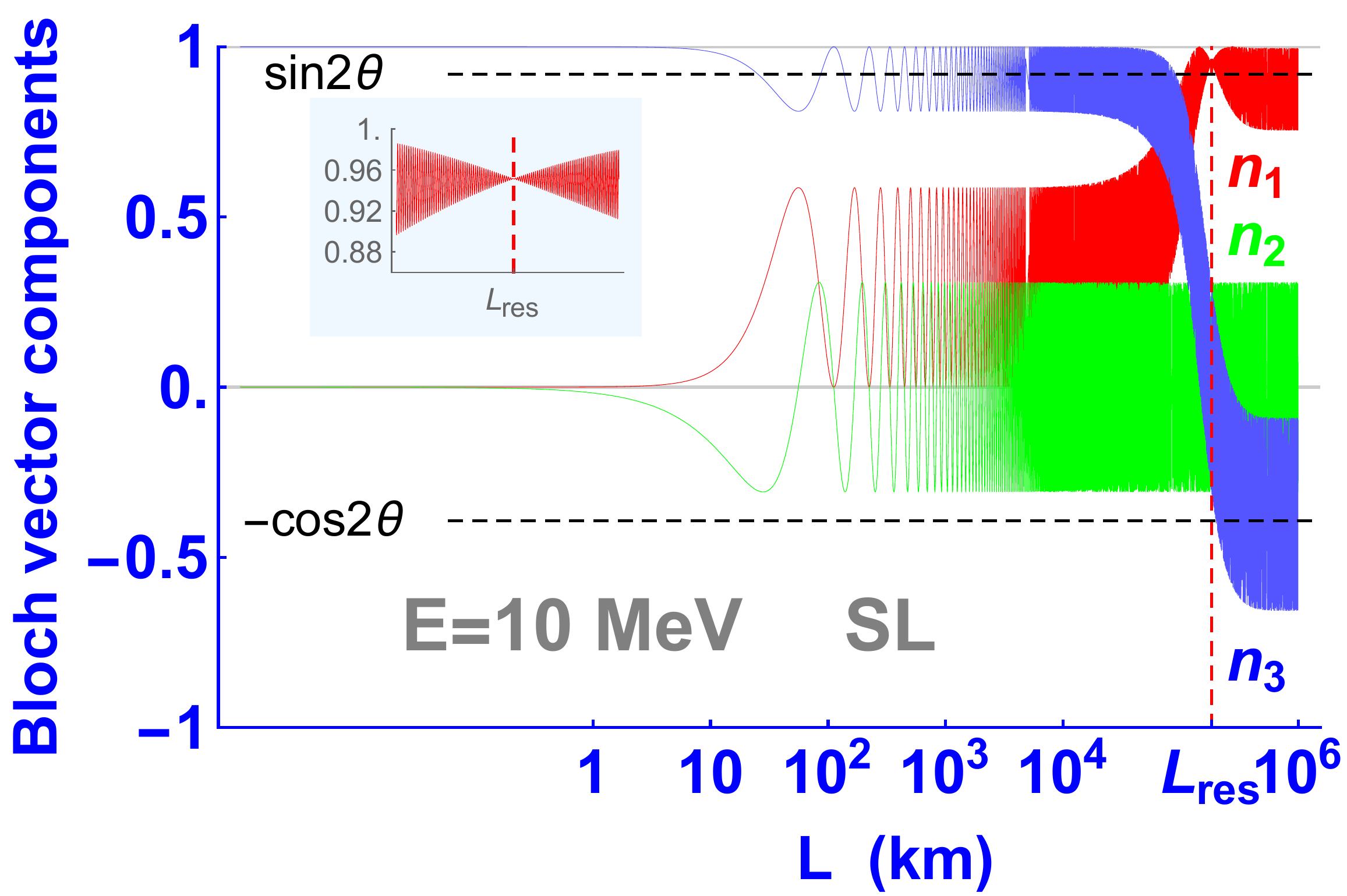}
    \caption{\label{fig:Fig1403d} Components of the Bloch vector  as a function of distance  from the centre of the Sun, $L$, for 5~MeV (left) and  10~MeV (right)  neutrino state,  according to the strictly linear (SL)  evolution equation~(\ref{eq:BlochEquationSL})  for two-flavour approximation with the $\nu_e$ in the initial state.
    The narrowing  observed  for the $n_1$ component at the  'resonance'  distance, $L_{\rm res}$, is shown in the inset. Dashed horizontal lines mark the values of components $n_1$ and $n_3$  corresponding to the pure state $\bm{\nu}_2=(\sin 2\theta,0,-\cos 2 \theta)$. }
    \end{center}
\end{figure}

Evolution  of the Bloch vector according to~(\ref{eq:BlochEquationSL}) from  the  $\nu_e$  initial state  of energy $E=5$~MeV and $E=10$~MeV  is  presented in Fig.~\ref{fig:Fig1403d}.
Each component oscillates  about a certain  locally averaged (over  one  oscillation length)  value  $\langle n_i \rangle$.
Mean values   of the components  $n_1$ and $n_3$  change  with distance in opposite directions  (recall that $\bm{n}(L)^2=1$) while
$n_2$   oscillates  near   zero, in accordance with the form of the  $\bm{\omega}_{\rm W}$ vector~(\ref{eq:VectorOmega1}).
Components   $n_1$ and $n_3$ manifest  a common flavour conversion phenomenon  in that they reach, as  the Wolfenstein potential goes to zero,  mean  values, $\langle n_1 \rangle$ and $\langle n_3 \rangle$, approximately corresponding  to the  $\nu_2$ mass eigenstate: $(\sin 2\theta,0,-\cos 2\theta)\approx (0.92,0,-0.39)$. We underline  that the asymptotic neutrino state is not  the  pure  mass state $\nu_2$   but  instead it is an oscillating state with the mean values of the components  approximately   corresponding to the $\nu_2$ state (a small deviation  is due to the finite energy). In consequence, the electron  neutrino survival probability, $p_{ee}(L)$,
is also an oscillating function of $L$ and it is effectively averaged  when measured on Earth, yielding
$\langle p_{ee}\rangle  \approx (1-\cos 2\theta)/2=\sin^2 \theta \approx 0.30$
(in the full three-flavour formalism one would expect a prediction  not by much higher,  matching well the measured value of $0.33\pm 0.02$ for  $^8$B  neutrinos~\cite{cite:Vissani1}).
The component $n_1$ of the Bloch vector has a  narrowing at the 'resonance' distance (a point-like  disappearance of oscillations)  which coincides with  the  mean third component, $\langle n_3 \rangle$, crossing zero. Thus the mean  $\nu_e$ survival probability   amounts to~0.5 at this point.  We also note that the initial $\nu_e$ does not achieve, even instantaneously, the  muon neutrino  state, $\nu_{\mu}$, given by $\bm{\nu}_{\mu}=(0,0,-1)$, anywhere during its evolution  through  the solar matter.

\begin{figure}[!htbp]
    \begin{center}
        \includegraphics[width=0.45\linewidth]{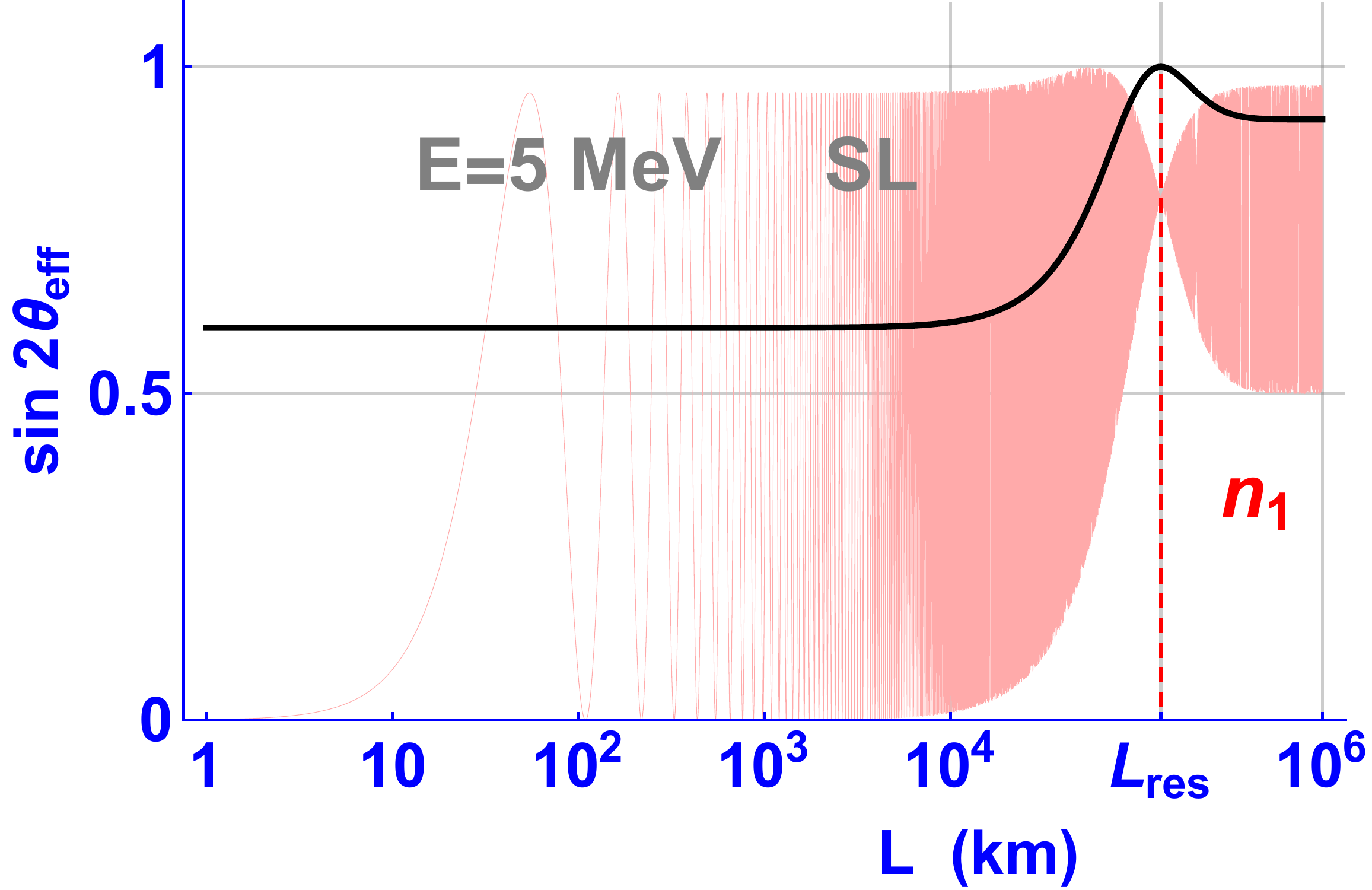}
        \hfill
        \includegraphics[width=0.45\linewidth]{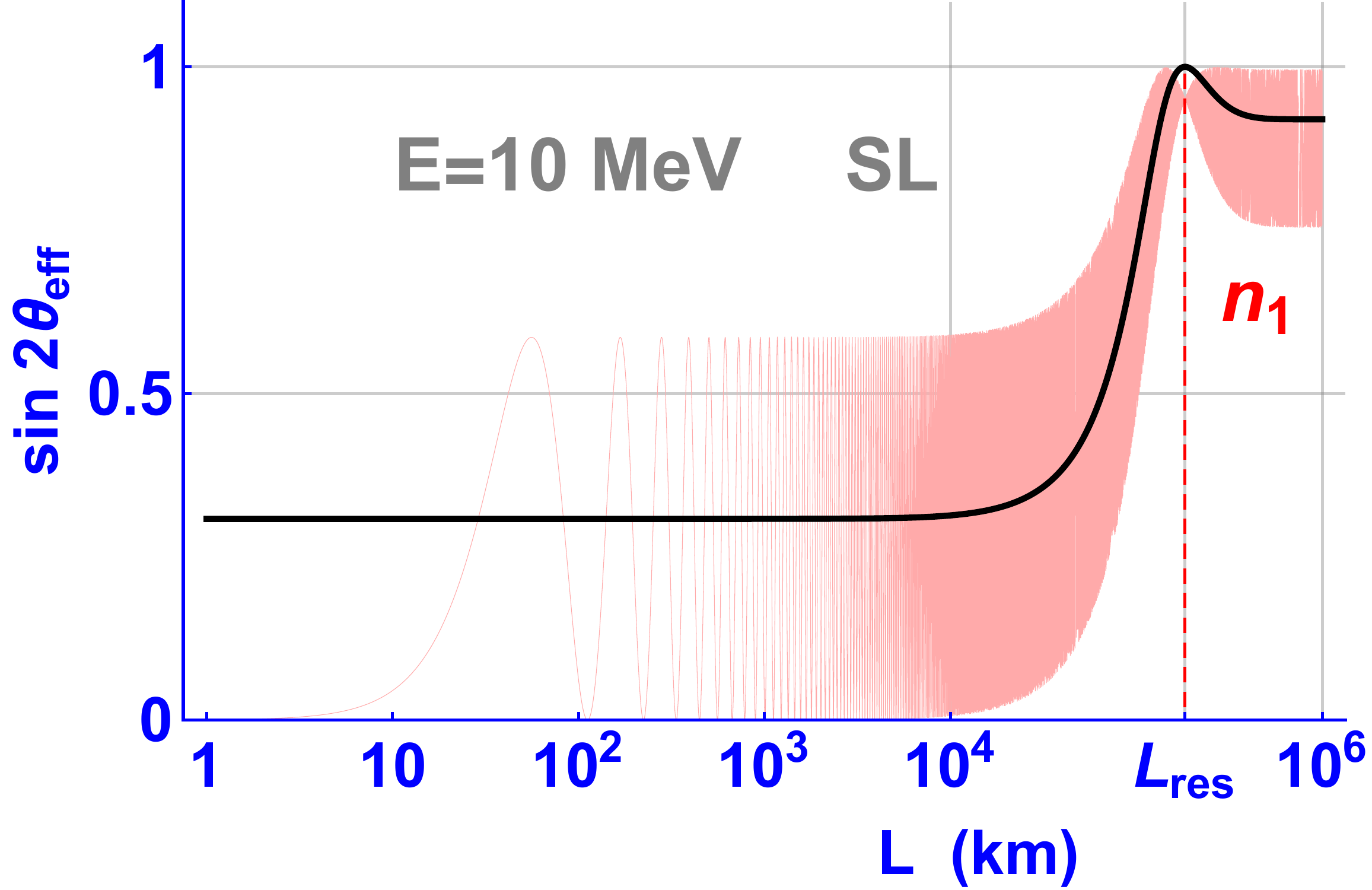}
    \caption{\label{fig:Fig1404ea}
    Effective mixing angle in matter, $\sin 2\theta_{\rm eff,W}$~(\ref{eq:EffectiveAngle2}), as a function of distance from the centre of the Sun, $L$, (solid line)  superimposed on the component $n_1$ of the Bloch vector   according to the strictly linear (SL)  evolution equation~(\ref{eq:BlochEquationSL})  for two-flavour approximation with $\nu_e$ in the initial state,  for 5~MeV (left) and  10~MeV (right)  neutrino state.}
    \end{center}
\end{figure}

Wolfenstein introduced two effective variables, reflecting matter effects  in neutrino oscillations, namely the   mixing angle, $\theta_{\rm eff,W}$, and the  mass-squared difference, $\Delta m_{\rm eff,W }^2$.
The  effective mixing  angle  reaches the value  $45^{\circ}$ at  the 'resonance' distance,  which  corresponds to  the  maximal mixing, i.e. $p_{ee}=p_{e\mu}=0.5$.
According to  the strictly   linear  formalism, and  using~(\ref{eq:EnergyGapL}) and~(\ref{eq:C1C2GapMin}), the effective mixing angle can be expressed  in terms of the energy gap as simply as
\begin{linenomath}\begin{equation}\label{eq:EffectiveAngle2}
\sin 2\theta_{\rm eff,W} (L) = \frac{d_{\rm W}(L_{\rm res})}{d_{\rm W}(L)}.
\end{equation}\end{linenomath}
Function~(\ref{eq:EffectiveAngle2}) and the evolution of the component $n_1$  of the Bloch vector  are  shown in Fig.~\ref{fig:Fig1404ea} for two energies, 5~MeV and 10~MeV. The effective mixing angle near the centre of the Sun  for $E=10$~MeV  yields  a few degrees only,  which indicates that   flavour mixing is  far much smaller than in the  vacuum (i.e., matter effects dominate).  Next,  the angle  attains  $45^{\circ}$ at the 'resonance' distance and then  asymptotically goes to the measured (vacuum)  value as  the Wolfenstein potential vanishes.
The higher energy, the more accurately   the  following relation is  approximated
\begin{linenomath}\begin{equation}\label{eq:EffectiveAngle3}
\langle n_1\rangle \approx \sin 2 \theta_{\rm eff,W}.
\end{equation}\end{linenomath}
The  effective mass-squared difference  can also  be  directly expressed  in terms of the energy gap
\begin{linenomath}\begin{equation}\label{eq:EffectiveMass2}
\Delta m_{\rm eff,W }^2 (L)=  2E \:d_{\rm W}(L)
\end{equation}\end{linenomath}
and one can show, using~(\ref{eq:C1C2GapMin}),  that $\Delta m_{\rm eff,W }^2$  goes to   $\Delta m^2$  as the  Wolfenstein potential vanishes
(the product $ \Delta m_{\rm eff,W }^2 \, \sin 2\theta_{\rm eff,W} = \Delta m^2 \, \sin 2\theta$ is energy independent).
Moreover, it can be numerically shown that  the effective oscillation length of the neutrino state  in  the solar  matter, $l_{\rm eff,W}$, again can be simply  expressed in terms of the energy gap
\begin{linenomath}\begin{equation}\label{eq:EffectiveLength1}
l_{\rm eff,W } (L)=  \frac{2 \pi }{ d_{\rm W}(L)}.
\end{equation}\end{linenomath}
The effective oscillation length  becomes that of free oscillations as  the Wolfenstein potential vanishes, $l_{\rm eff,W} \rightarrow 2\pi / \omega_0$, where $\omega_0=\Delta m^2/2E$.
Relations~(\ref{eq:EffectiveAngle2}), (\ref{eq:EffectiveMass2}) and~(\ref{eq:EffectiveLength1}) suggest  that  the effective mixing of  neutrino flavour  states and the character of oscillations  in  the solar matter is actually governed by the energy gap, i.e., the distance of the system  from the structural  instability point,
while   its minimal value  sets a scale for the   effective mixing angle.
This   interpretation illustrates  the   advantage  of  our  approach to a deeper understanding of the  dynamics  underlying the linear evolution of neutrino flavour oscillations in the solar matter.

\subsection{Quasilinear evolution of solar neutrinos}\label{sec:Damping}

Now, we proceed to  describe  evolution of the neutrino state  in the solar matter  according to  the quasilinear  evolution equation~(\ref{eq:vonNeumann2}).
The  subsystem-environment interaction is ascribed to the additional   generator, $G$, appearing in~Eq.\ref{eq:vonNeumann2}, hence it is justified to adopt only  the free  Hamiltonian  in this case, $H_{\rm f}$ (cf. Appendix~I)

\begin{linenomath}
\begin{gather}\label{eq:Wolfenstein0}
H_{\rm f} = E\cdot I + \frac{\Delta m^2}{4E}
\begin{pmatrix}
- \cos 2\theta \;\;\; &  \sin 2\theta\\
\sin 2\theta \;\;\;& \cos 2\theta
\end{pmatrix}.
\end{gather}
\end{linenomath}

Vector $\bm{\omega}=\bm{\omega}_0$ then  simply reads
\begin{linenomath}\begin{equation}\label{eq:VectorOmega3}
\bm{\omega}_0=\left (\frac{\Delta m^2 }{2E}\sin 2\theta,\:0,\:-\frac{\Delta m^2}{2E}\cos 2\theta \right).
\end{equation}\end{linenomath}
Since the subsystem-environment interaction arises  through the electron number density in the solar plasma, it is natural  to contain it  in the   vector $\bm{g}$
\begin{linenomath}\begin{equation}\label{eq:VectorG2}
\bm{g}(L) = g_e(L) \bm{e},
\end{equation}\end{linenomath}
where  $g_e(L)=\sqrt{2}{\rm G_F} n_e(L)$ and $\bm{e}$ is a unit vector.
Note that  the term "potential" is rather inadequate in reference to this case.
It is convenient to  express the scalar product $\bm{\omega}_0 \bm{g}$  with the aid of  a new  angle $\sigma$
\begin{linenomath}\begin{equation}\label{eq:AngleSigma}
\bm{\omega}_0 \bm{g}= \frac{\Delta m^2}{ 2E} \:g_e \cos\sigma
\end{equation}\end{linenomath}
that leads to the following parametrisation of the unit vector

\begin{linenomath}
\begin{eqnarray}
\bm{e}=
\begin{pmatrix}
\tfrac{\cos\sigma+\cos\eta \cos 2\theta }{ \sin 2\theta}\\
\pm  \tfrac{ \sqrt{-\Big( \cos (\eta+2\theta)+\cos\sigma \Big) \Big(\cos(\eta - 2\theta)+\cos\sigma \Big) } }{ \sin 2\theta }\\
\cos\eta
\end{pmatrix}.
\label{eq:UnitVector2}
\end{eqnarray}
\end{linenomath}

\begin{figure}[!htbp]
    \begin{center}
        \includegraphics[width= 12.0cm]{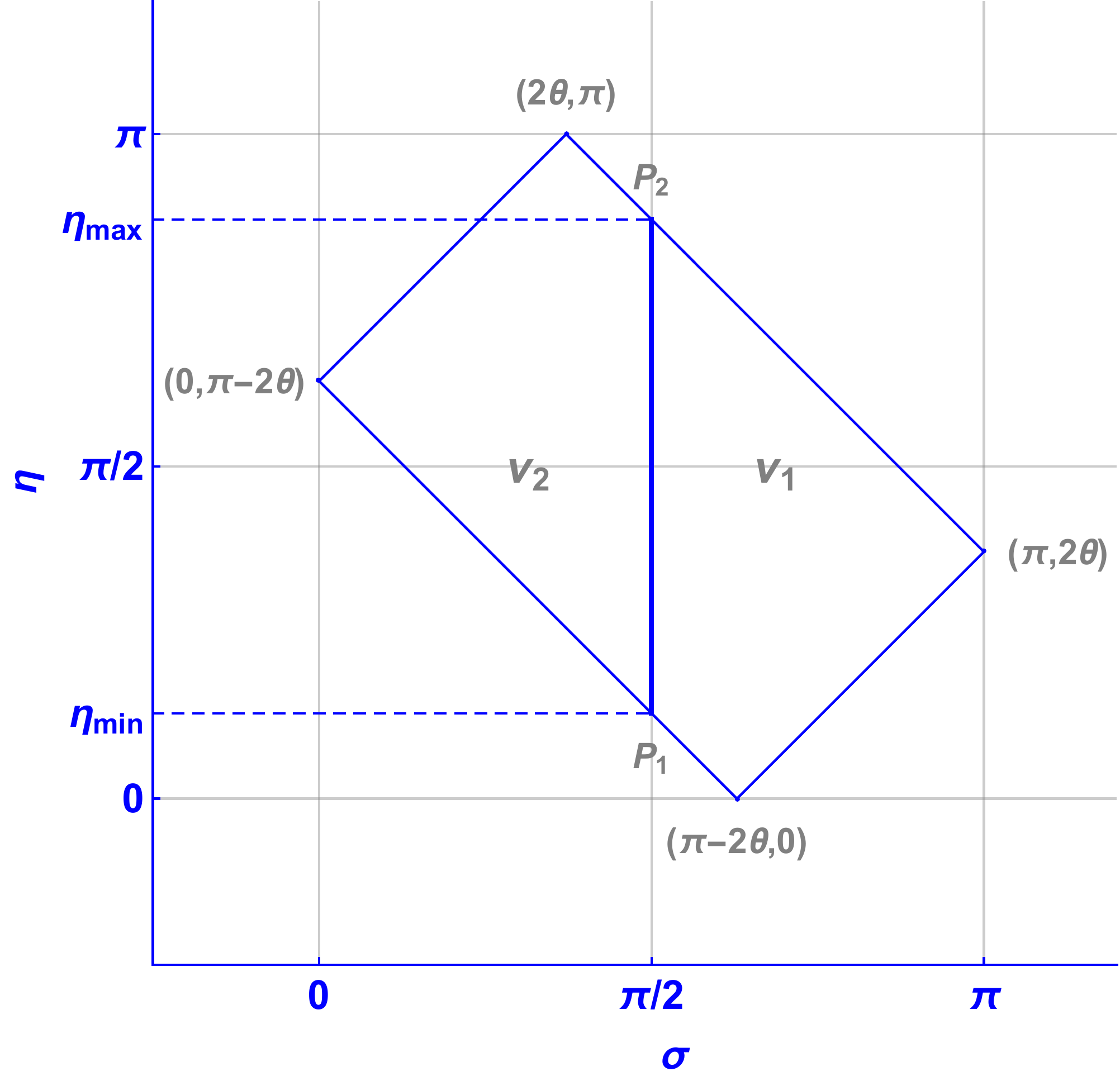}
    \caption{\label{fig:Fig1205Regions} Rectangular space of parameters ($\sigma,\eta$), following from~(\ref{eq:UnitVector2}),~(\ref{eq:EtaMinMax1}).
    The thick vertical line marks the physical range of $\eta$ at  $\sigma=\pi /2$ (transition line); coordinates of the vertices   are shown in brackets.}
    \end{center}
\end{figure}
Positive-definiteness of the expression under the square-root in~(\ref{eq:UnitVector2})
fixes the rectangular space of parameters for   $\eta$  as shown in Fig.~\ref{fig:Fig1205Regions} in the range $[0,\pi]$.
The vertical   line at $\sigma=\pi /2$, corresponding to the perpendicular orientation of  the vectors  $\bm{\omega}_0$ and $\bm{g}$,   distinguishes two extreme  values of $\eta$ located  at the boundary of the physical region
\begin{linenomath}
\begin{equation}\label{eq:EtaMinMax1}
\eta_{\rm min}=\frac{\pi}{2}-2\theta; \;\;\;\;\;
\eta_{\rm max}=\frac{\pi}{2}+2\theta
\end{equation}
\end{linenomath}
(points $P_1$ and $P_2$ in  Fig.~\ref{fig:Fig1205Regions}).
The angles  $\sigma$ and $\eta$  play an analogous  role in the parametrisation of $\bm{g}$, i.e., fix  its orientation  in the lepton flavour space,
like  the mixing angle $\theta$  does  for the parametrisation of $\bm{\omega}$. We note however  that the angle $\sigma$ arises only in matter, i.e., when $\bm{g}\ne 0$~(\ref{eq:AngleSigma}),  and  its vacuum value is physically  undefined.
Now, in view  of~(\ref{eq:UnitVector2}), the  invariants $C_1$ and $C_2$~(\ref{eq:C1C2}) take the following form
\begin{linenomath}\begin{equation}\label{eq:c1c2G}
C_{\rm 1,QL}(L) =  \left ( \frac{\Delta m^2}{2E}\right ) ^2 - g_e^2(L) \;\;\;\;\; C_{\rm 2,QL}(L) = \frac{\Delta m^2}{2E} g_e(L)  \cos\sigma (L),
\end{equation}\end{linenomath}
and the  corresponding  energy gap~(\ref{eq:EnergyGap1})   reads
\begin{linenomath}\begin{equation}\label{eq:EnergyGapQL}
d_{\rm QL}(L) = \Bigg [ \;    \Bigg (   \Bigg(\frac{\Delta m^2}{2E}\Bigg)^2 - g_e(L)^2 \Bigg )^2 + 4 \Bigg (\frac{\Delta m^2}{2E} g_e(L) \cos\sigma (L) \Bigg )^2   \;   \Bigg ]^{\frac{1}{4}}.
\end{equation}\end{linenomath}
Invariant $C_1$ vanishes for $L=L_{\rm ins}$  which fulfils the following condition
\begin{linenomath}\begin{equation}\label{eq:InsCondition1}
g_e(L_{\rm ins})=\frac{\Delta m^2}{2E}.
\end{equation}\end{linenomath}
The relationship of $L_{\rm ins}$  and  $L_{\rm res}$ follows from~(\ref{eq:ResCondition1}) and~(\ref{eq:InsCondition1}). Since  $V_e (L_{\rm res})/g_e (L_{\rm ins}) = \cos 2\theta$ and  $n_e (L)$ is a decreasing function of $L$,  the following relation holds: $L_{\rm ins} <  L_{\rm res}$.
For example, for $E=10$~MeV, $L_{\rm ins}\approx 118800$~km < $L_{\rm res}\approx 182800$~km.
Invariant $C_2$ vanishes  independently for   $\sigma=\pi/2$.
The threshold energy in this description  results from the condition $C_1(0)=0$,
\begin{linenomath}\begin{equation}\label{eq:ThresholdEnergy2}
E_{\rm th,QL} =  \frac{\Delta m^2}{2 g_e(0)} \approx 3.09~{\rm MeV}.
\end{equation}\end{linenomath}

The strictly perpendicular orientation of vectors  $\bm{\omega}_0$ and $ \bm{g}$ in the solar matter can be excluded on the experimental basis.  Indeed, if  $\sigma=\pi/2$,
the oscillating neutrino state reaches the point $L=L_{\rm ins}$ at  which  both invariants vanish,  $C_1=C_2=0$, and so does the energy gap,
$d_{\rm QL}(L_{\rm ins})=0$~(\ref{eq:EnergyGapQL}).  One then  has to do with  maximal  instability of the system  caused by the structural instability point,  in consequence of which  the character of the evolution drastically changes, which manifests as follows.
From the moment of neutrino creation  at $L=0$, the  oscillations are fully suppressed (see below) until the system reaches the  structural instability point
at  which  the evolution acquires a form of  free  oscillations  for  all three components of the Bloch vector.  In particular the  mean value of the third component of the Bloch vector   amounts to zero, $\langle n_3\rangle = 0$,  corresponding to the survival probability  $p_{ee}=0.5$, independently of energy (above threshold), in gross disagreement  with  the experimental results for $10$~MeV  $^8$B  neutrinos.
Thus  the evolving state of the system  must omit   the  structural instability point     by adopting a  nonzero value of the invariant $C_{2,QL}~($\ref{eq:c1c2G}) which can be achieved if the  value of the angle $\sigma$   differs  from $\pi /2$.
The transition  line  at $\sigma=\pi /2$ separates  two regions on the $(\sigma,\eta)$ plane, denoted $\nu_1$ and $\nu_2$ in Fig.~\ref{fig:Fig1205Regions}. If $\sigma <\pi /2$ the initial state  $\bm{\nu}_e$  asymptotically converges to    $\bm{\nu}_2$, as indicated   the  measurements on Earth,  while would do  to   $\bm{\nu}_1$  in   the opposite case. Such pattern is reminiscent of the existence of two separated phases, $\nu_1$ and $\nu_2$.
Now, the key observation is  that only when the value   of $\sigma $ is contained within  a  narrow  band
about the transition line,  approximately  $10^{-2}$~rad wide,
an energy-dependent prediction for the  $\nu_e$ survival probability would result, in agreement with   the  measurements.
We thus introduce   a dimensionless  perturbation  parameter, $\delta <0$, to describe the deviation of the vectors  $\bm{\omega}$ and $\bm{g}$  from perpendicularity. The effective angle $\sigma$ in the solar matter is then expressed as follows
\begin{linenomath}\begin{equation}\label{eq:SigmaDelta1}
\sigma = \frac{\pi}{2} + \delta
\end{equation}\end{linenomath}
(as $\delta$ should  vanish  in the vacuum,   the angle $\sigma$  accordingly would become  $\pi /2$  which can be taken as a reference, not the vacuum value). Following~(\ref{eq:SigmaDelta1}), the Casimir invariant $C_2$~(\ref{eq:c1c2G}) can be written as  $\bm{\omega}\bm{g}\approx |\bm{\omega}|  |\bm{g}|\,\delta $ which is a scalar under rotations in the flavour space. Bearing in mind Fig.~\ref{fig:Fig1205Regions}
we note that if the value of $\eta$ is set to $\eta_{\rm max}$, only the minus sign in~(\ref{eq:SigmaDelta1}) is admissible
since the plus sign  would move the value of $\sigma$   beyond  the physical  region.
Nevertheless, as separate  studies have shown, the exact value of $\eta$ is not relevant for the current considerations.

In order to find a plausible formula for $\delta$,  we argue along the following line.
A   nonzero  value of  $\delta$    arises due to  a  non-vanishing  electron number density, except  where it is constant and  the instability point cannot be reached (e.g., terrestrial matter).  This justifies assuming    $\delta$  proportional to the  gradient  of the radial distribution of the electron number density~(\ref{eq:InvMorse1}),  $\delta  \propto h^{\prime}_e <0 $.
We  adopt an effective parametrisation  of  $\delta$ in the solar  energy range in a factorised form, as  a product of three terms, $\delta (E,L) = C \,F(E) \, h^{\prime}_e(L)$, where $C$ is  a constant factor  of  dimension  ${\rm km}$ in order   to account  for the unit ${\rm km^{-1}}$  of the  gradient, and $F(E)$ -- a  dimensionless function of the neutrino energy.
The  partial product of the first two terms, defined as  $C_F(E)=C\,F(E)$, will also be used below.
We assume the  coefficient $C$ to  be expressed in terms of the universal constants with an  additional  multiplicative  factor $f$, in general depending on the mixing angle,  serving to fulfill  the  criteria  specified below. The function $F(E)$ is  expressed in terms  of a  dimensionless ratio  involving the  quantities describing the system, i.e., $g_e(0)/\omega_0$~($\omega_0=\Delta m^2 /2E$).
\begin{figure}[!htbp]
    \begin{center}
        \includegraphics[width=0.65\linewidth]{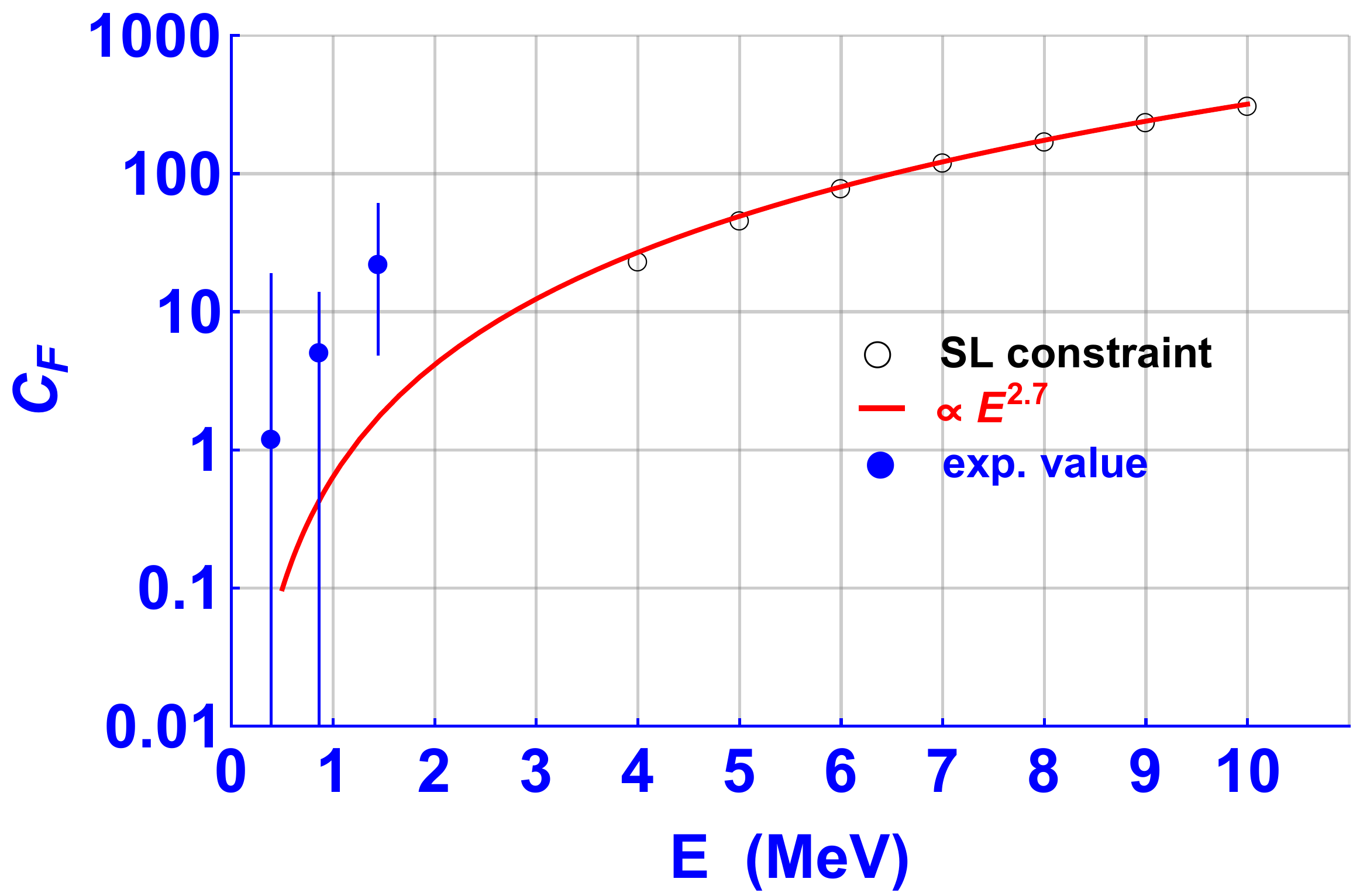}
    \caption{\label{fig:Fig2101a} Empirical parametrisation of the function $C_F\propto E^{2.7}$~(\ref{eq:Delta1}),  determined by matching  the $\delta$-dependent QL   prediction for survival probability, $p_{ee}$, to points (open circles) representing the $p_{ee}$ values at discrete energies for the SL evolution (solid line); full points  with error bars  represent  values of $C_F$ matching the experimental survival probability  for pp, $^7$Be and pep   neutrinos. }
    \end{center}
\end{figure}
We parameterise  the function $F(E)$, i.e., the energy dependence of $\delta$, in terms of a power  $r$ of the ratio $g_e(0)/\omega_0$,  yielding  the following explicit formula
\begin{linenomath}\begin{equation}\label{eq:Delta1}
        \delta (E,L) = f\,\frac{\hbar c}{g(0)} \left ( \frac{g(0)}{\omega_0} \right )^r  \; h^{\prime}_e(L),
\end{equation}\end{linenomath}
where the  parameter $r$ remains to be determined; note also that  $g(0)/\omega_0 = E/ E_{\rm th,QL}$, where  the threshold energy  is given by~(\ref{eq:ThresholdEnergy2}).
If accurate  measurements of  the neutrino  survival probability were available in a range  of  energies, the value of the parameter $r$ could  be determined  experimentally. Meanwhile there is only one such  useful constraint coming from  the  measurement at  $E=10$~MeV for  $^8$B neutrinos   while those  for pp, $^7$Be and pep  neutrinos are of a limited  value  due to  their very  large uncertainties.
We thus tentatively  chose to determine an  indicative  value  of $r$,   obtained
by demanding that  certain  predictions of the  QL evolutions agree with those of the SL evolution;  in particular,    the mean  asymptotic (as the electron number density  becomes negligible) survival probability as a function of the neutrino energy, $p_{ee}(E)$ and its asymptotic  oscillation amplitude (depth). It  appeared that these criteria  entailed   also a  good agreement of  two other asymptotic quantities  introduced hereunder  -- energy transfer and speed of the  state evolution~(Sec.~\ref{sec:EnergyFlow} and \ref{sec:Speed}).
In order to determine the  parameter $r$ and the coefficient $f$   we  computed several values of  $C_F (E)$ at discrete energies above the threshold ($4<E<10$~MeV)   that lead to QL predictions   matching  those of  the SL  evolution, as detailed above ("SL constraint").
We  fitted the resulting values with a function proportional to $E\,^r$, yielding $r\approx 2.7$,  as  shown in Fig.~\ref{fig:Fig2101a}. An agreement of the depths of oscillations at asymptotic distances   for the SL and QL evolutions can be  achieved for  $f\approx 0.82$.
These particular  values of $r$ and $f$  arise  as a consequence of adopting  the input values for  $\Delta m^2$, $\theta$  and $g(0)$ specified earlier.

Aside from  the above effective  parametrisation, obtained relatively  to the SL evolution,  we independently determined the  value of $C_F$   for each  individual  measured survival probability at low energies  (pp, $^7$Be and pep neutrinos) in such a way that the QL prediction reproduces each  experimental value, with  uncertainties  of $C_F$ resulting from  matching the experimental values  altered  by the corresponding  experimental uncertainty.  These results  are also shown Fig.~\ref{fig:Fig2101a} as experimental points with error bars and  suggest that the $C_F$, i.e.,  $\delta$,   is compatible below the threshold with the prediction following from~(\ref{eq:Delta1}).

\begin{figure}[!htbp]
    \begin{center}
        \includegraphics[width=0.45\linewidth]{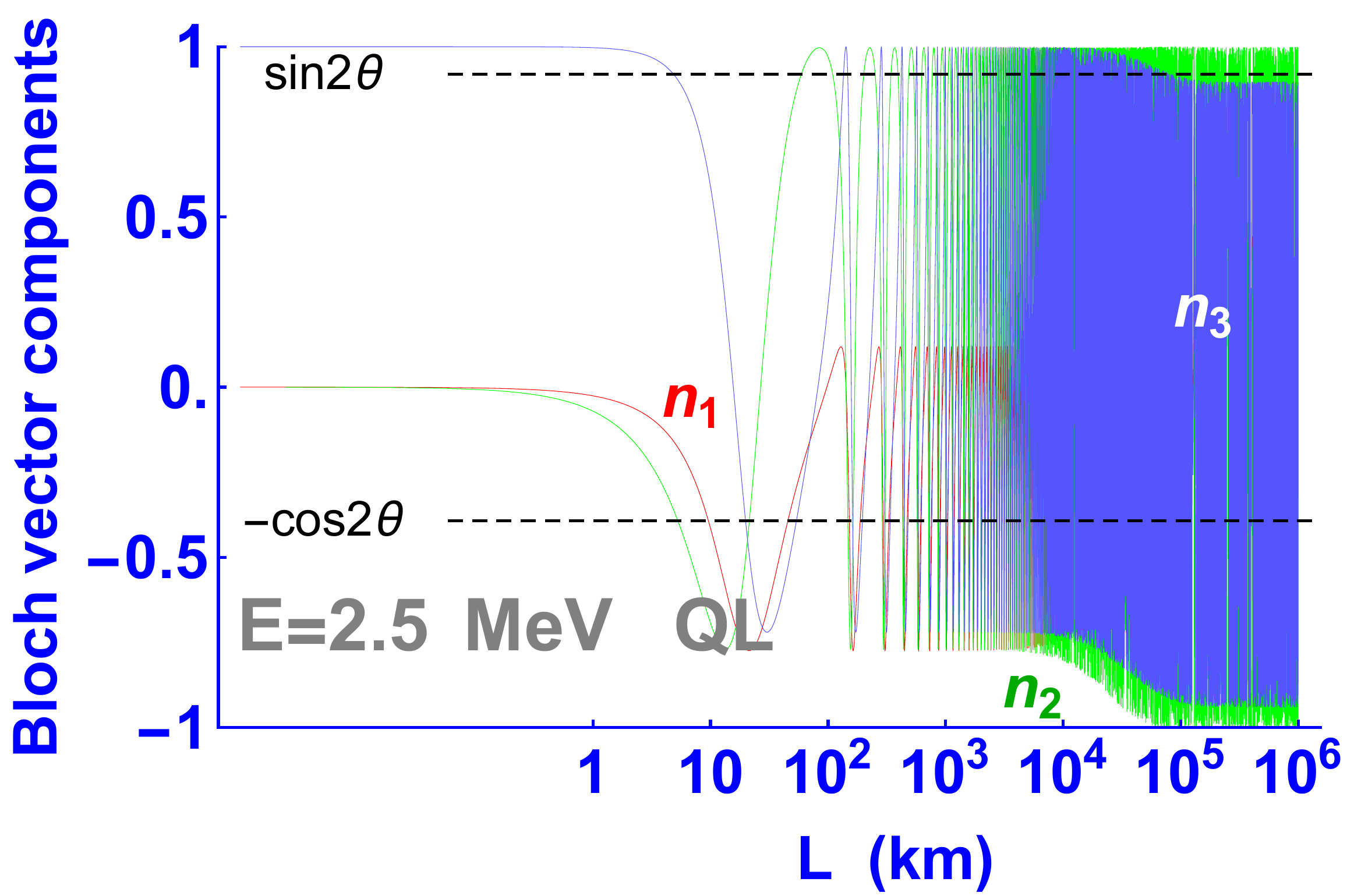}
        \hfill
        \includegraphics[width=0.45\linewidth]{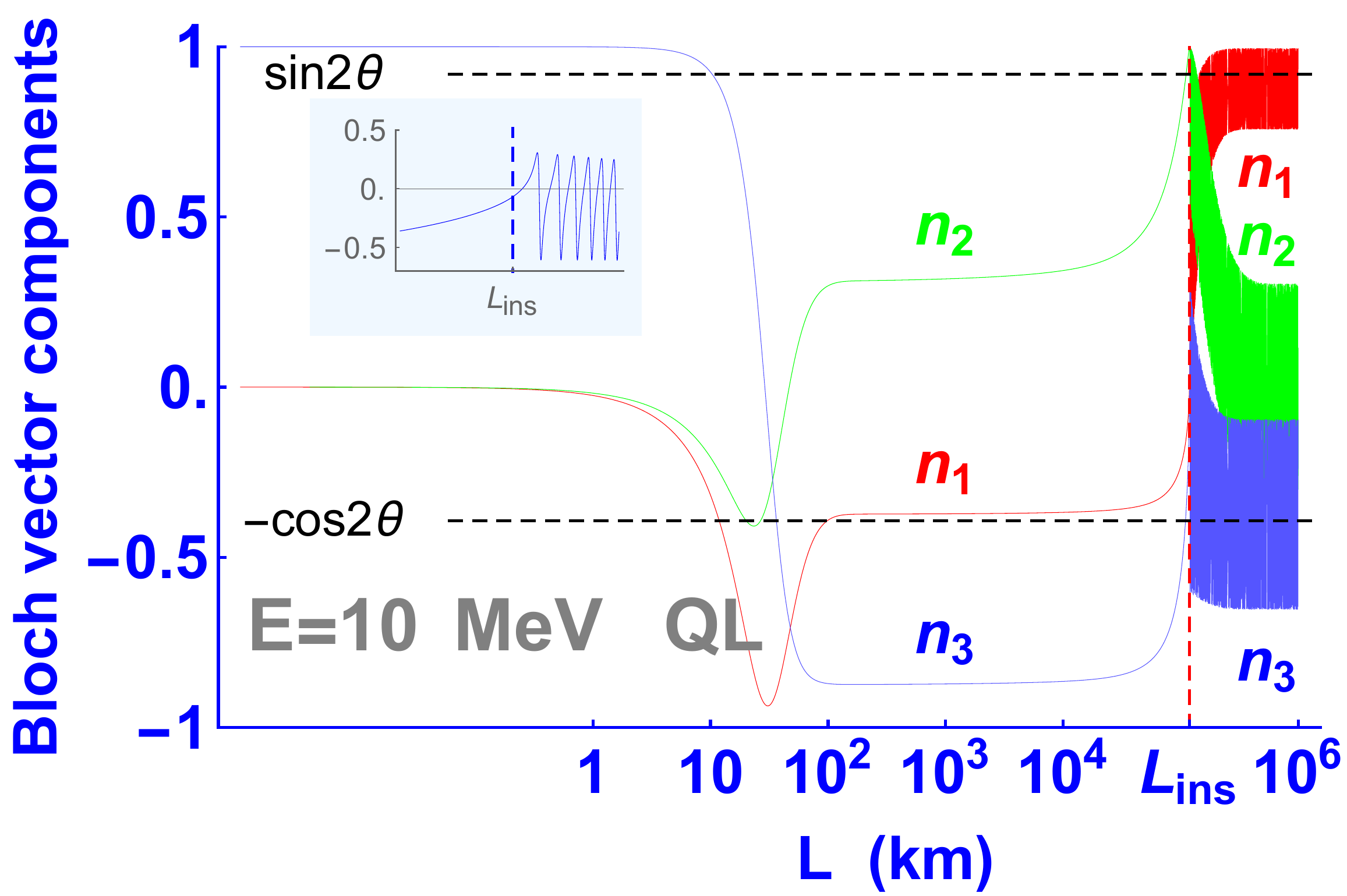}
    \caption{\label{fig:Fig1403e} Components of the Bloch vector  as a function of distance  from the centre of the Sun, $L$, for a sub-threshold 2.5~MeV (left) and  10~MeV (right)  neutrino state,  according to the quasilinear evolution equation~(\ref{eq:BlochEquation}) with $\eta=\eta_{\rm max}$   and $\delta$  given by~(\ref{eq:Delta1}) and detailed  in the text,   for two-flavour approximation with the $\nu_e$ in the initial state.  Details of the component $n_3$ about the instability point,  $L_{\rm ins}$,   are shown in the inset for $E=10$~MeV. Dashed horizontal lines mark the values of components $n_1$ and $n_3$  corresponding to the pure state $\bm{\nu}_2=(\sin 2\theta,0,-\cos 2 \theta)$. }
    \end{center}
\end{figure}

Evolution of the Bloch vector   with $\delta$  given by~(\ref{eq:Delta1}) and  $r=2.7$  is presented in Fig.~\ref{fig:Fig1403e} where  the solution of~(\ref{eq:BlochEquation})  is shown for a sub-threshold energy  $E=2.5$~MeV and for~$E=10$~MeV  ($\eta=\eta_{\rm max}$) and   $\nu_e$ in the initial state.
In the former case the evolution has a  periodic but nonlinear  character. For energies above the threshold,   oscillations of all three components of the Bloch vector are suppressed  inside the Sun  and  appear only as  the system  reaches  the instability distance, $L_{\rm ins}$, and subsequently  converge  to a constant mean value, similarly as in the strictly  linear case.  The influence  of the  structural instability point on the evolution of the system for energies above threshold is manifested exactly by  this sudden change of the character of the evolution, from non-oscillatory to oscillatory,  shown in the inset for the component $n_3$.
Again  the asymptotic neutrino state is not  the  pure  mass state $\nu_2$   but   an oscillating state with the mean values of the components  approximately   corresponding to  $\nu_2$.  The  electron  neutrino survival probability, $p_{ee}(L)$,
is also an oscillating function of $L$ as the electron number density vanishes.
The component $n_3$ of the Bloch vector  crosses zero at the instability distance and  the mean  $\nu_e$ survival probability   amounts to~0.5 at this point (maximal mixing). Neither in this case the initial $\nu_e$  achieves the  muon neutrino  state  during its evolution.

\subsection{Comparison of linear and quasilinear evolutions}

\begin{figure}[!htbp]
    \begin{center}
        \includegraphics[width=0.75\linewidth]{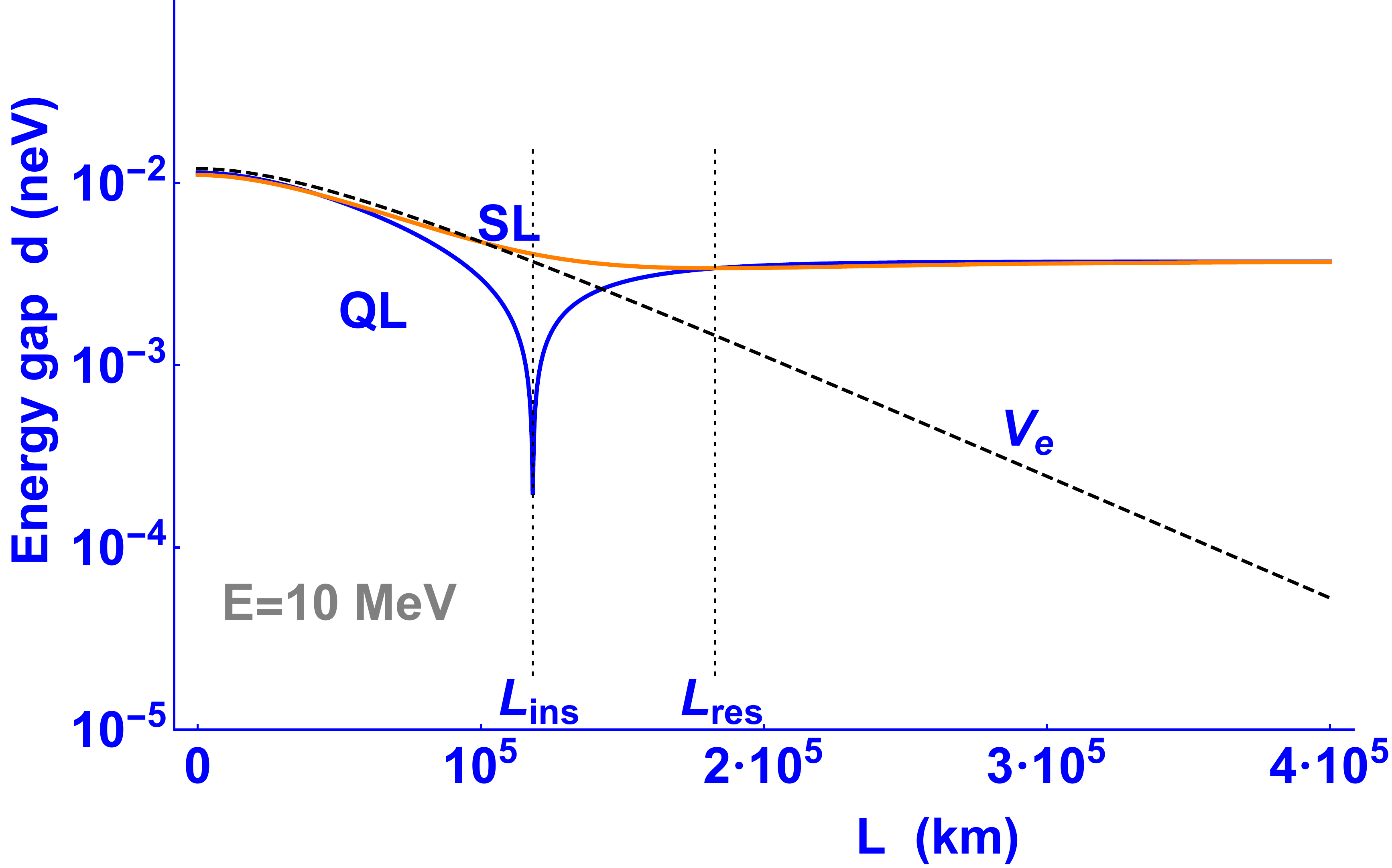}
    \caption{\label{fig:Fig1402l} Energy gap~(\ref{eq:EnergyGap1}) for  strictly linear  (SL) and quasilinear (QL) evolution of 10~MeV  neutrino  state; note the shallow minimum at $L_{\rm res}$ in the SL  case. The Wolfenstein potential, $V_e$, is plotted on the same scale,  marked by  dashed line.}
    \end{center}
\end{figure}

\begin{figure}
\centering
\includegraphics[width=0.45\linewidth]{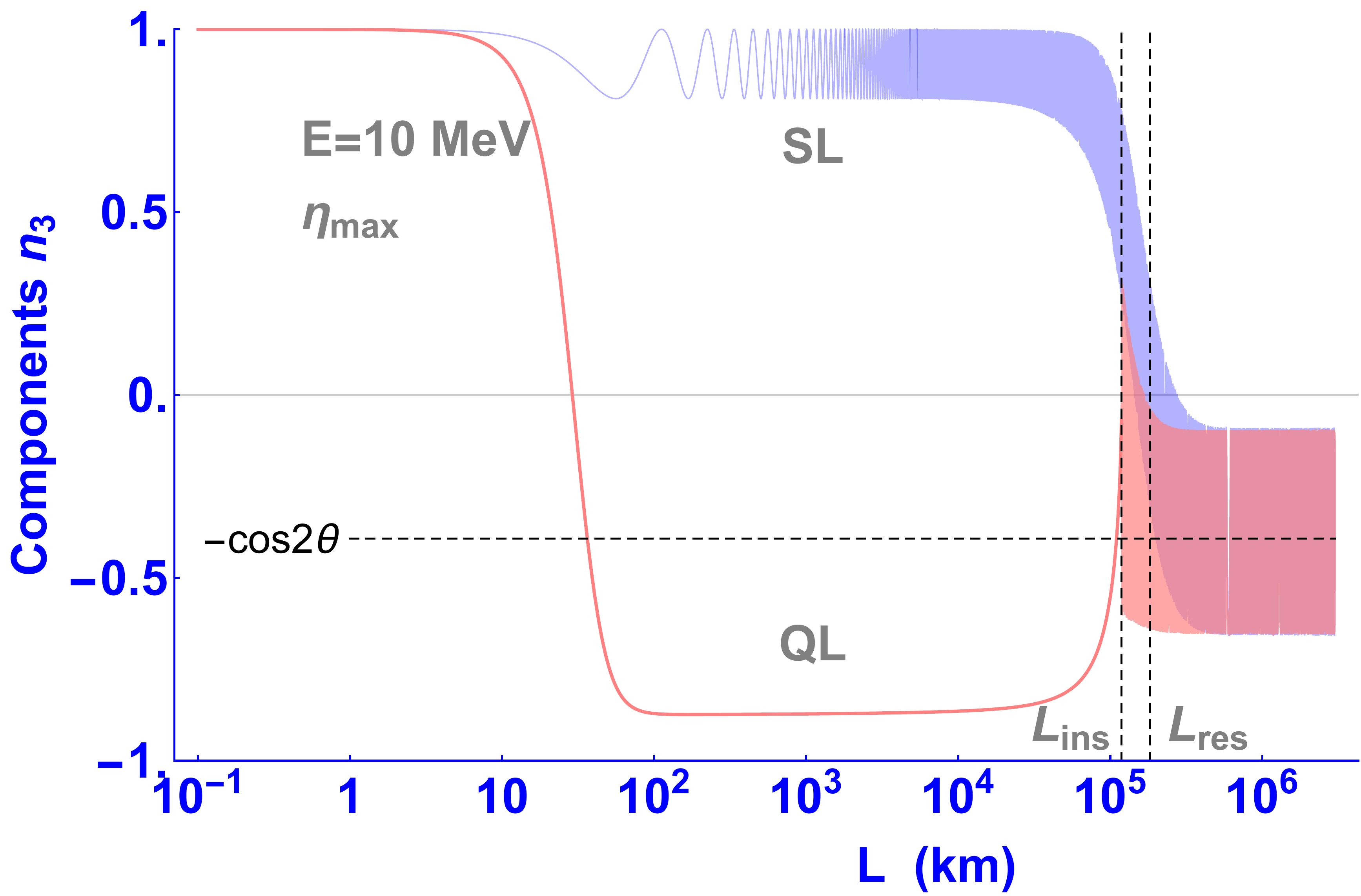}
\hfill
\includegraphics[width=0.45\linewidth]{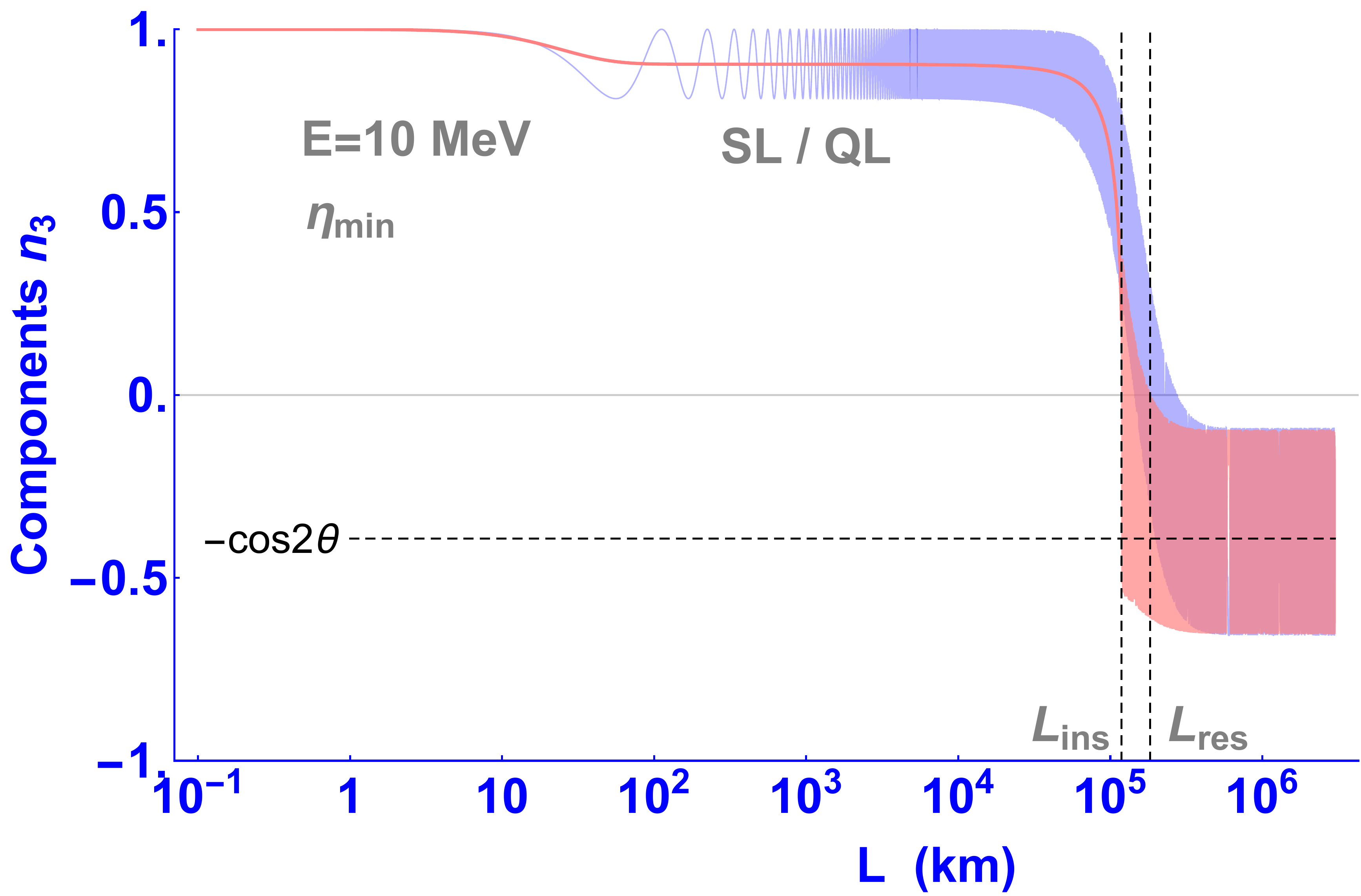}
\caption{\label{fig:Fig1403comp10MeV} Comparison of the components $n_3$  of the Bloch vector  as  functions of distance from the centre of the Sun, $L$, for
 a 10~MeV  neutrino state according to the strictly  linear (SL)~(\ref{eq:BlochEquationSL})
   and   quasilinear (QL)~(\ref{eq:BlochEquation})  evolution  with  $r=2.7$ for  $\eta=\eta_{\rm max}$ (left) and  $\eta\approx\eta_{\rm min}$ (right).
Distances of the structural instability point, $L_{\rm ins}$,
and  the  'resonance'  point, $L_{\rm res}$,   are marked by dashed  vertical lines. Dashed horizontal line marks the $n_3$ component of the Bloch vector corresponding to the pure state $\bm{\nu}_2=(\sin 2\theta,0,-\cos 2 \theta)$.}
\end{figure}

Since the $L$-dependent  energy gap, i.e.,  distance of the system from the structural instability point,  plays a key role in the evolution of state in  both strictly linear and quasilinear descriptions,   in  Fig.~\ref{fig:Fig1402l} we compare the respective functions,  given by equations~(\ref{eq:EnergyGapL}) and (\ref{eq:EnergyGapQL}).
In the strictly linear  case,  the function  $d_{\rm W}(L)$  exhibits a shallow minimum at   $L=L_{\rm res}$. In  the quasilinear case, the function $d_{\rm QL}(L)$  is characterised by a deep and sharp (but differentiable) minimum  at  $L=L_{\rm ins}$.  Both functions have approximately the same  values at $L=L_{\rm res}$  and   converge to one common asymptotic $L$-dependence.
Predictions for the third component of the Bloch vector for the linear and the quasilinear ($\eta=\eta_{\rm max}$) neutrino evolutions  are compared in Fig.~\ref{fig:Fig1403comp10MeV} for $\eta=\eta_{\rm max}$ and $\eta=\eta_{\rm min}+0.02$ (the value of $\eta$ near the minimum  increased by a tiny amount to allow the angle $\sigma=\pi /2+\delta$ remain within the physical region).  Mean values and depths of  survival probabilities for both  cases  in the asymptotic limit can be brought to agreement  by adjusting only the   factor $f$  of the perturbation parameter $\delta$~(\ref{eq:Delta1}).   The two asymptotically  oscillating functions are relatively shifted in phase (not resolved on the figure), however of no   effect on the  measurements on Earth.

\begin{figure}[!htbp]
    \begin{center}
        \includegraphics[width=0.8\linewidth]{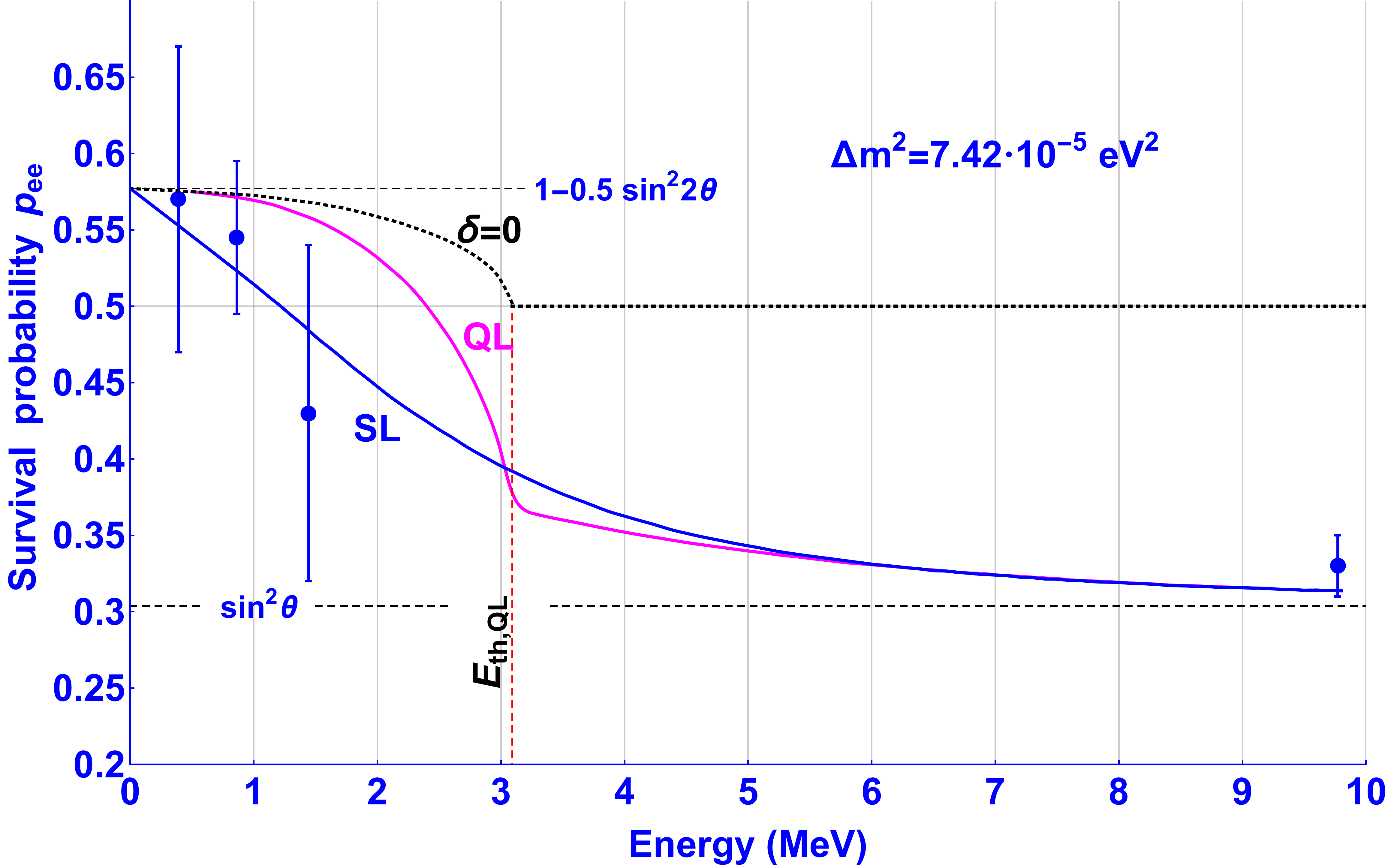}
    \caption{\label{fig:Fig1710a} Predicted energy dependence of the neutrino  survival probability, $p_{ee}$, for the strictly linear (SL)  and quasilinear (QL) with $r=2.7$  ($\eta = \eta_{\rm max}$)     evolution computed using  $\Delta m^2=7.42\cdot 10^{-5}$~eV$^2$.
     Dashed horizontal lines mark  the values for $E\rightarrow 0$ and $E\rightarrow \infty$, amounting to    $p_{ee}=1-\frac{1}{2}\sin^2 2\theta\approx 0.58$   and $p_{ee}=\sin^2 \theta\approx 0.30$, respectively. Dashed vertical line indicates the threshold energy for the QL evolution, $E_{\rm th,QL}$, amounting to  $3.09$~MeV~(\ref{eq:ThresholdEnergy2}). Prediction for $\delta = 0$ is also shown (dotted line).}
    \end{center}
\end{figure}

Since  measurements  on Earth  regard  only the mean value of the asymptotic survival probability, an important information is  delivered by its energy dependence, which is shown in Fig.~\ref{fig:Fig1710a}.
The  energy dependence is monotonic  in the strictly linear (SL)  description and reaches the asymptotic value of approx. 0.3 for $E=10$~MeV.
In the quasilinear (QL)  description,   using $\eta=\eta_{\rm max}$ and $\delta$  given by~(\ref{eq:Delta1}) with $r=2.7$, the energy dependence is distinctly  different. One can observe   an "ankle"  at the threshold energy  beyond which the survival probability  tends   towards an agreement with  the strictly  linear  evolution (as was assumed). Adopting $\eta\approx\eta_{\rm min}$ for the QL evolution only insignificantly alters the predictions  below the threshold.
The "ankle" in the quasilinear evolution the threshold energy denotes a change in the energy dependence dynamics resulting from  the  transition  between two contrasting  evolution modes:   oscillatory  and non-oscillatory (cf. Fig.~\ref{fig:Fig1403e}). The prediction for $\delta = 0$ is also shown.
Comparing any model with  predictions  of the linear formalism  could be conclusive   only if  the precision of the data about the  threshold, $1\div 4$~MeV,  was  significantly improved. This is   very difficult to achieve since  the neutrino flux  is low  in that energy range (mainly  the pep and  ${\rm ^7Be}$ channels) in   the presence of  a  significant radioactive background.

\section{Neutrino in solar environment}

\subsection{Neutrino--Sun energy transfer}\label{sec:EnergyFlow}

\begin{figure}[!htbp]
    \begin{center}
        \includegraphics[width=0.8\linewidth]{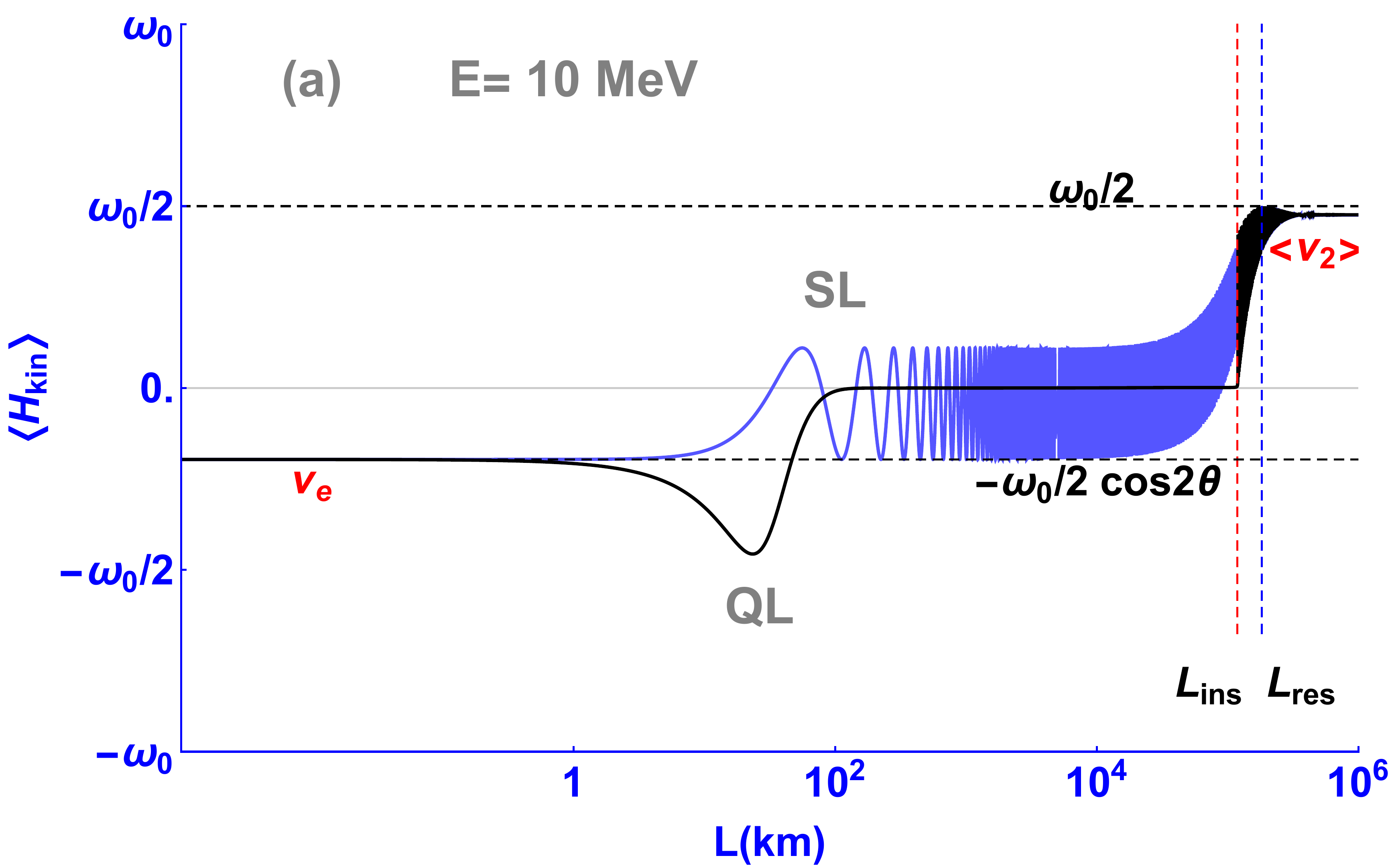}\\
         \includegraphics[width=0.8\linewidth]{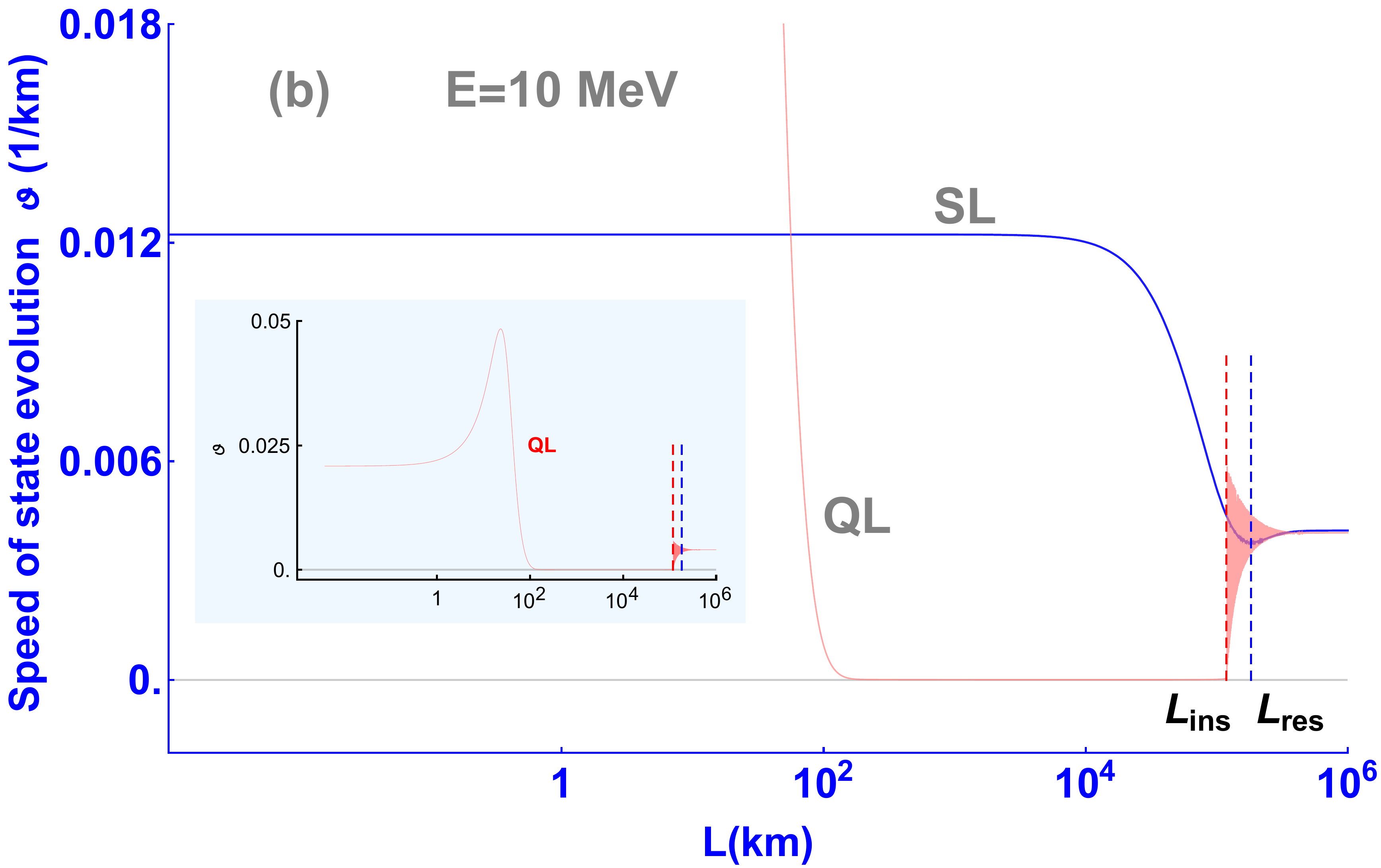}
    \caption{\label{fig:Fig1400Double10MeV} (a) Mean  kinetic  Hamiltonian, $\langle H_{\rm kin} \rangle$, in units of $\omega_0=\Delta m^2/2E$, as a function of the distance from the centre of the Sun, $L$, according to the strictly linear (SL)  and quasilinear  (QL) ($\eta = \eta_{\rm max}$)  evolution in the solar matter for
  $E=10$~MeV  neutrino state; $\nu_e$ and  $\langle \nu_2 \rangle$  denote the initial and the final states, the  latter referring to  its oscillating character with the mean Bloch vector corresponding approximately  to that of the mass eigenstate $\nu_2$;   (b) speed  of the strictly  linear (SL)   and quasilinear (QL)  state evolution~(\ref{eq:Speed0}) (km$^{-1}$); the  full function for the QL evolution is shown  on a separate scale in the inset.}
    \end{center}
\end{figure}

Given  the Bloch vector, one can address a natural  question in the context of  open systems --  mean  energy transfer between the propagating neutrino   and the Sun, intriguing despite an unmeasurable  quantity involved,  of the order of  $\Delta m^2 / 4 E$.
The mean value of the kinetic part of the free  Hamiltonian, $\langle H_{\rm kin} \rangle$~(cf. Appendix~I), which corresponds to a difference of "kinetic terms"  for  mass eigenstates $m_1$ and $m_2$, is given by the standard  quantum-mechanical formula
\begin{linenomath}\begin{equation}\label{eq:Averages1}
\langle H_{\rm kin} \rangle =\mathrm{Tr}\hspace{2pt}(H_{\rm kin}\:\rho )   = \frac{1}{2}\bm{\omega}_0\bm{n},
\end{equation}\end{linenomath}
where $\rho$  is the  density matrix, $\bm{\omega}_0$ is given by~(\ref{eq:VectorOmega3})  and $\bm{n}=\bm{n} (L)$ is the Bloch vector obtained by solving~(\ref{eq:BlochEquationSL}) or~(\ref{eq:BlochEquation}), depending on the considered case.
A comparison of both   mean kinetic terms of the free   Hamiltonians as a function of $L$ for the linear and quasilinear propagation  is  shown  in Fig.~\ref{fig:Fig1400Double10MeV}a  for $E=10$~MeV.
Although functional dependencies differ  inside the Sun, the average values in  both descriptions reach the same  limit
as  the electron number density  vanishes, provided  the asymptotic oscillation depths agree  (independently  of the parameter $\eta$ in the QL description).
The  value of  $\langle H_{\rm kin} \rangle$ at the  $\nu_e$ creation point, $L=0$, is negative and amounts to $\langle  H_{\rm kin}\rangle  = -(\Delta m^2/4E) \cos 2\theta$ while it reaches an almost   asymptotic  value of  $\Delta m^2/4E$  as  the neutrino state  becomes   the mean $\nu_2$ state.
This finding indicates  that the net gain of kinetic energy of  the neutrino system is  positive by an amount $(\Delta m^2/4E)(1+ \cos 2\theta)$
after the state has made its way through the solar matter.
According to the SL  evolution, the  increase occurs monotonically; in  the QL  description  it is sharp and arises suddenly at the instability point  $L=L_{\rm ins}$.  The  fact that the  propagating neutrino  and the Sun constitute  an open system  is   more pronounced   in the quasilinear description of the neutrino  evolution.

\subsection{Speed of the state evolution}\label{sec:Speed}

In  quantum information theory an important notion   is  the  speed of the state evolution, denoted by $\vartheta$. Among different definitions~\cite{cite:Wigner1,cite:Anandan1,cite:Brody2,cite:Campaioli1} (cf. Appendix~II) the most natural and effective concept involves  the time derivative of the Bloch vector which, adapted to the present formalism of two-flavour oscillations,  leads to the following expression  with time replaced by  $L$ as the evolution parameter
\begin{linenomath}
\begin{equation}\label{eq:Speed0}
\vartheta (L) = \sqrt{\frac{1}{2} \bm{n}^{\,\prime}(L)^2}.
\end{equation}
\end{linenomath}
Using~(\ref{eq:BlochEquation}) yields
\begin{linenomath}
\begin{equation}\label{eq:Speed2}
\vartheta (L)  = \tfrac{1}{\sqrt{2}} \sqrt{\bm{\omega}^2\bm{n}^2 -(\bm{\omega}\bm{n})^2  +\bm{g}^2  +(\bm{n}^2-2) (\bm{g}\bm{n})^2 +2(\bm{g}\times\bm{\omega})\bm{n} },
\end{equation}
\end{linenomath}
which for pure states ($\bm{n}^2=1$) reduces to
\begin{linenomath}
\begin{equation}\label{eq:Speed3}
\vartheta_{\rm pure} (L)  = \tfrac{1}{\sqrt{2}} \sqrt{\bm{\omega}^2 -(\bm{\omega}\bm{n})^2  +\bm{g}^2  - (\bm{g}\bm{n})^2 +2(\bm{g}\times\bm{\omega})\bm{n} },
\end{equation}
\end{linenomath}
leading to   a simple formula  in the case of the  linear   evolution ($\bm{g}=0$)
\begin{linenomath}
\begin{equation}\label{eq:Speed4}
\vartheta_{\rm W} (L)  = \tfrac{1}{\sqrt{2}} \sqrt{\bm{\omega}^2 -(\bm{\omega}\bm{n})^2   }
\end{equation}
\end{linenomath}
with $\bm{\omega}$ given by~(\ref{eq:VectorOmega1}) and $\bm{n}$  being the  solution of~(\ref{eq:BlochEquationSL}) with the  initial condition
$\bm{n}(0)=\bm{\nu}_e$.
An analogous,  though more complicated, formula can be  obtained from (\ref{eq:Speed3}) for the quasilinear evolution, with the aid of~(\ref{eq:VectorOmega3}) and (\ref{eq:VectorG2}).
The function~(\ref{eq:Speed4}),  shown in Fig.~\ref{fig:Fig1400Double10MeV}b for $E=10$~MeV, has a minimum  very near  $L=L_{\rm res}$ and a certain asymptotic value.  Thus for finite energies  a minimal value of the evolution speed   occurs  near the 'resonance' distance. However it can be  checked using~(\ref{eq:Speed4}) that the  value of the evolution speed  would become zero with vanishing electron number density, should  the asymptotic neutrino state   become  strictly $\nu_2$. This would be the case   for asymptotic energies when  the outgoing   neutrino state  becomes  $\bm{\nu}_2=(\sin 2\theta,0,-\cos 2\theta)$. Meanwhile,  also the minimum would disappear as  the evolution speed would decrease and remain  zero beyond  the 'resonance' distance. On the other hand,  the higher energy, the better does  the asymptotic neutrino  state approximate the $\nu_2$ state.

The speed of the state evolution and the effective oscillation length in the linear picture~(\ref{eq:EffectiveLength1}) are numerically connected  as follows
\begin{linenomath}
\begin{equation}\label{eq:SpeedOscLength}
l_{\rm eff,W}(L) = \frac{\sqrt{2}}{\vartheta_{\rm W} (L)}
\end{equation}
\end{linenomath}
and thus the relation
\begin{linenomath}
\begin{equation}\label{eq:SpeedGap1}
\vartheta_{\rm W} (L) = \frac{d_{\rm W} (L)}{\sqrt{2} \pi}
\end{equation}
\end{linenomath}
repeatedly  supports the interpretation that the speed of the evolution of state is driven by the energy gap, $d_{\rm W}(L)$.

\section{Summary, discussion and conclusions}

We presented a new  description of neutrino propagation  in the solar matter  starting from the assumption  that the whole is  an open system due to the weak  interactions.   In consequence we exploited  the  von Neumann equation modified   by way of  a quasilinear extension,  allowed by quantum mechanics for open systems~\cite{cite:RC1}. In the resulting evolution  equation~(\ref{eq:BlochEquation}) for the Bloch vector of the neutrino state, $\bm{n}(L)$,   interactions of neutrinos  were  described  through the  Hamiltonian, $H$,  and an additional generator, $G$, containing  information about  the  neutrino interactions with the solar environment.  We   discussed the  solutions for the Bloch vector in the context of  $\sg{SL(2,C)}$ covariance group of this system. The Casimir invariants $C_1$ and $C_2$ of  $\sg{SL(2,C)}$ determine the energy gap, $d(L)$~(\ref{eq:EnergyGap1}), i.e., an interval in the space of invariants, characterising  structural instability and  'resonance' points of equation~(\ref{eq:BlochEquation}) as well as the magnitude of the influence of the structural instability point on the evolution of the system.
It was shown that the structural instability point should be  identified with the configuration $(C_1,C_2)=(0,0)$ while the 'resonance' point is obtained  by minimising the energy gap~(\ref{eq:EnergyGapL}) (Fig.~\ref{fig:Fig1402l}).

We first applied this extended   formalism  to  the  strictly linear (SL)  von Neumann   equation   with the Hamiltonian~(\ref{eq:WolfensteinH}) containing the interaction part specified by  the Wolfenstein potential. We solved  Eq.~\ref{eq:BlochEquationSL}  in the two-flavour approximation, with the initial condition corresponding to the electron-neutrino.    The SL von Neumann  equation is a special case of the quasilinear equation~(\ref{eq:BlochEquation}) with $G=0$.  We  obtained  the  Bloch vector comprising the complete  information about the evolution of the system during propagation in   the solar matter.
The  change of the  locally averaged survival probability near the 'resonance' point (Fig.~\ref{fig:Fig1403d}) can be  interpreted   as due to the influence of the structural  instability point.

Next we focused on solutions of the modified von Neumann equation~(\ref{eq:BlochEquation}) in which terms involving the additional generator $G$ were preserved, resulting in its quasilinearity. We restricted ourselves to the case of a free Hamiltonian~(\ref{eq:Wolfenstein0}) with the  neutrino interactions with the environment   contained solely in the generator $G$ via the flavour vector $\bm{g}_e(L)$ (\ref{eq:VectorG2}). The principal feature of the  QL evolution described by~(\ref{eq:BlochEquation}) under the above  choice is such that the survival probability inside the Sun,  $L < L_{\rm ins}$, is a non-oscillating function of $L$ which can be interpreted, according to the classical nomenclature, as characteristic for a strongly damped (overdamped) system. In this case the  structural instability point can be reached during evolution of the system,  $C_1(L_{\rm ins})=C_2(L_{\rm ins})=0$, when the  flavour vectors $\bm{\omega}$ and $\bm{g}$ are perpendicular. In order to make  the system   bypass  this point it  is necessary to introduce an energy and $L$-dependent perturbation, $\delta$, to  the direction of the vector $\bm{g}$,   fixing the orientation of the generator $G$ in the flavour space. We proposed an \textit{ansatz} for $\delta$~(\ref{eq:Delta1}) which assures almost the same $\nu_e$  survival probability at the highest measured energy point (10~MeV)  as in the SL case.
The differences between the SL and QL approaches  may be possibly observed in the  detailed energy dependence for  neutrino energies below  and about the threshold, i.e.,  in the energy range $1\div 4$~MeV~(Fig.~\ref{fig:Fig1710a}). However, obtaining corresponding measurements of the neutrino survival probability  with small uncertainties is challenging.  We underline that in both, SL and QL, cases the neutrinos leaving the  Sun are not in the pure  $\nu_2$ state but in a state for which the locally averaged  $\nu_e$ survival probability  is nearly the same as for $\nu_2$. This can be seen  from the Fig.~\ref{fig:Fig1403d} and  Fig.~\ref{fig:Fig1403e} where the evolution of the neutrino state is presented and oscillations in  the asymptotic state are evident.

We have shown that  a distinguished  distance, appearing   in the SL  description, denoted $L_{\rm res}$,  corresponds  to a minimum of the energy gap or a maximum of the effective mixing angle~(Fig.~\ref{fig:Fig1404ea}), however   assigning a resonant character to these  aspects of evolution is too far reaching. As regards the   "adiabatic flavour conversion", indicated in the literature on the MSW effect,  the formalism  based on the  density  matrix does not require introducing such  concepts. The initial state $\nu_e$ evolves throughout the entire trajectory of the neutrino state in the Sun  and   ends up in an oscillating state with mean values of the Bloch vector components corresponding to the state $\nu_2$.

We have also dealt with two new  related issues.  One effect, although interesting  but rather unmeasurable,  is    energy   transfer  in the neutrino flux  on the way through  the  solar matter.  From Fig.~\ref{fig:Fig1400Double10MeV}(a) it follows that during propagation  inside the Sun the neutrino gains  'kinetic' energy, i.e.,  energy  is transferred to it from the environment, irrespectively of the SL or QL character of the evolution. This observation  supports  the open system scenario worked out  in this paper. However, the  gap  between the outgoing neutrino mean energy and the maximal energy which corresponds to the state $\nu_2$ supports the earlier conclusion that the outgoing state   is near  but not identical with  $\nu_2$.
The other  study  regarded the speed of the neutrino state evolution. Indeed,  one can see  from Fig.~\ref{fig:Fig1400Double10MeV}(b)  that the asymptotic speed of the outgoing state is different from zero which  should be the case for the pure state $\nu_2$.  Each of the  above  quantities,  obtained from  the strictly linear and quasilinear equations,  respectively,  are  in agreement in the limit  of the vanishing  solar electron number  density.

As a general conclusion, we have  demonstrated that the evolution of the neutrino state in the solar matter, aside from the MSW formalism,  can be also  described  in terms of a   quasilinear evolution equation,  resulting from treating the neutrino and the Sun as an open system. Both  approaches  lead to similar results for the survival probability on Earth, with predictions  notably different   inside the Sun. Thus for obvious reasons it  cannot be resolved by a direct measurement  which one is proper in the solar scenario. A hope to achieve this in  an  indirect way,  through the  energy dependence of the survival probability, is rather faint. One can however search for  signatures of the  quasilinear evolution    in  the past and future  terrestrial  experiments with  known baselines and beams passing through the Earth's mantle.

\section{Appendices}

\subsection{Appendix I}
Under the assumption of constancy of the neutrino momentum,  neutrino  kinetic energies $E_i = \sqrt{p^2 + m_i^2}$ are eigenvalues of the free Hamiltonian, $H_{\rm free}$,  which takes the following  form in the flavour basis (without approximations)
\begin{equation}\label{eq:Appendix1a0}
H_{\rm free} = E\cdot I + H_{\rm kin}
\end{equation}
where $I$ is the identity matrix,   $E$ is the mean energy, i.e., $ E = \tfrac{1}{2} (E_1 + E_2)$ and $H_{\rm kin}$ denotes the "kinetic" part of the free Hamiltonian

\begin{eqnarray}\label{eq:Appendix1a}
H_{\rm kin} =  \frac{\Delta m^2}{4E}
\begin{pmatrix}
- \cos 2\theta \hspace{5mm} &  \sin 2\theta\\
\sin 2\theta \hspace{5mm} &  \cos 2\theta
\end{pmatrix}.
\end{eqnarray}

The full Hamiltonian includes the interaction term,
$H_{\rm int}$, which can be rewritten in the  flavour basis  as follows

\begin{eqnarray}\label{eq:Appendix1b}
H_{\rm int}=
\begin{pmatrix}
 V_e(L)  \hspace{5mm} &  0\\  0 \hspace{5mm}&  0  \end{pmatrix} = \frac{V_e(L)}{2}\cdot I + \frac{1}{2}\begin{pmatrix}  V_e(L)  \hspace{3mm}  &  0\\  0 \hspace{3mm}&  -V_e(L)
\end{pmatrix}.
\end{eqnarray}

Thus the full Hamiltonian, $H_{\rm tot}=H_{\rm free} + H_{\rm int}$,    reads
\begin{multline}
H_{\rm tot}(L) = \left ( E  +\frac{V_e (L)}{2} \right)\cdot I  \\
+ \frac{1}{2}
\begin{pmatrix}
- \tfrac{\Delta m^2}{2E} \cos 2\theta  +V_e(L) &  \sin 2\theta\\
\sin 2\theta  &  \tfrac{\Delta m^2}{2E} \cos 2\theta - V_e(L)
\end{pmatrix},
\label{eq:Appendix1c}
\end{multline}
where the second term is identified as  the Wolfenstein Hamiltonian.
The  term proportional to the unit matrix in~(\ref{eq:Appendix1c}) drops out from the von Neumann equation~(\ref{eq:vonNeumann1}) as a result of commutation properties of the unit matrix  leading to the traceless MSW Hamiltonian~(\ref{eq:WolfensteinH}).
For any  value of $L$, neutrino energy varies between the eigenvalues  $E_{\pm}(L)$ of the Hamiltonian~(\ref{eq:Appendix1c})
\begin{linenomath}\begin{equation}\label{eq:Appendix1d}
E_{\pm}(L) = \left ( E+\frac{V_e(L)}{2} \right ) \pm \frac{1}{2} \sqrt{\left( \frac{\Delta m^2}{2E}\sin 2\theta \right )^2 + \left( V_e(L)- \frac{\Delta m^2}{2E}\cos 2\theta \right )^2 },
\end{equation}\end{linenomath}
where the square root term, cf.~(\ref{eq:EigenValuesH}), determines  the limits of the energy range and   can be written as $\pm \frac{1}{2} \sqrt{C_1(L)}$~(\ref{eq:C1C2MSW}).
As the  Wolfenstein  potential vanishes,  the  energy range~(\ref{eq:Appendix1d}) reaches
\begin{linenomath}\begin{equation}\label{eq:Appendix1e}
E_{\pm} = E  \pm  \frac{\Delta m^2}{4E}.
\end{equation}\end{linenomath}
In a general case  the eigenvalues  of the  generator~(\ref{eq:HiG})
\begin{linenomath}\begin{equation}\label{eq:Appendix1f}
H+i G = \frac{1}{2} (\bm{\omega}+ i \bm{g})\bm{\sigma}
\end{equation}\end{linenomath}
are complex and take the form
\begin{linenomath}\begin{equation}\label{eq:Appendix1g}
\lambda_{\pm} = \pm \frac{1}{2} \sqrt{(\bm{\omega}^2-\bm{g}^2)+2i\bm{\omega}\bm{g}} = \pm \frac{1}{2}\sqrt{C_1(L) + 2i C_2(L)}\equiv \pm \lambda (L)
\end{equation}\end{linenomath}
while the energy gap~(\ref{eq:EnergyGap1}) can be written as
\begin{linenomath}\begin{equation}\label{eq:Appendix1h}
d(L) =  \left ( (\bm{\omega}^2-\bm{g}^2)^2 + 4 (\bm{\omega}\bm{g})^2 \right )^{\frac{1}{4}} = \left( C_1(L)^2+4 C_2(L)^2 \right )^{\frac{1}{4}} = 2|\lambda_{\pm}|
\end{equation}\end{linenomath}
and  then  $E_+ - E_- = d(L)$ at a given energy, $E$.
We also note that, since ${\rm Tr}\rho=1$,   any term proportional to the unit matrix in an expansion of the generator $G$ drops out from the modified von Neumann equation too.

\subsection{Appendix II}
The most natural and effective definition of the speed of the state evolution   is the following
\begin{linenomath}
\begin{equation}\label{eq:SpeedA0}
\vartheta (t) = \sqrt{\mathrm{Tr}(\partial_t \rho)^2}.
\end{equation}
\end{linenomath}
In the case of  ${\rm n}$-dimensional Hilbert space the density matrix, $\rho$, can be  represented as
\begin{linenomath}
\begin{equation}\label{eq:SpeedA1}
\rho = \frac{1}{\mathrm{n}}\Bigg ( I + \sum_{\alpha=1}^{\mathrm{n}^2-1}  n_{\alpha} \lambda_{\alpha}   \Bigg ),
\end{equation}
\end{linenomath}
where $\lambda_{\alpha}$ are the generalised Hermitian, traceless Gell-Mann matrices with  $\mathrm{Tr}(\lambda_{\alpha}\lambda_{\beta})=\delta_{\alpha\beta}$.  The components $n_{\alpha}$ of the generalized, ${\rm n}^2-1$-dimensional, Bloch vector $\bm{n}$, are real  and    $\bm{n}^2=\sum_{\alpha=1}^{{}\rm n^2-1}  n_{\alpha} \leq 1$,   which constitutes the Euclidean norm in the convex space  of Bloch vectors.  By means of~(\ref{eq:SpeedA1}), the speed~(\ref{eq:SpeedA0}) takes  the  form
\begin{linenomath}
\begin{equation}\label{eq:SpeedA2}
\vartheta = \sqrt{\frac{1}{\mathrm{n}}  \dot{\bm{n}} ^2 }.
\end{equation}
\end{linenomath}
Adapted to the solar neutrino case of $L$-dependent density matrix,  $\rho(L)$, evolution speed $\vartheta(L)$, with the distance $L$ as the evolution parameter, can be expressed as
\begin{linenomath}
\begin{equation}\label{eq:SpeedA3}
\vartheta (L) = \sqrt{\frac{1}{\mathrm{n}}  \bm{n}^{\,\prime}(L)^2 }.
\end{equation}
\end{linenomath}
Now, by means of~(\ref{eq:vonNeumann2}), equation~(\ref{eq:SpeedA3}) can be written as
\begin{linenomath}
\begin{multline}\label{eq:SpeedA4}
\vartheta (L) = \sqrt{\mathrm{Tr}\Big(\partial_L \rho (L)\Big)^2} =\\
\sqrt{2\Bigg( \mathrm{Tr} \Big(( G^2 + H^2) \rho ^2\Big) +\mathrm{Tr}(G\rho)^2 - \mathrm{Tr}(H\rho)^2 + i \mathrm{Tr}([H,G]\rho^2) -2 \mathrm{Tr} (G\rho^2)^2 \Bigg)
}
\end{multline}
\end{linenomath}
(agreeing  units requires accounting for the $\hbar c$ term on the r.h.s. of the above and the following equalities).
In particular, for a pure state, $\rho^2 = \rho$,  (\ref{eq:SpeedA4}) can be simplified
\begin{linenomath}\label{eq:SpeedA5}
\begin{multline}
\vartheta_{\rm pure} (L) = \sqrt{\mathrm{Tr}\Big(\partial_L \rho (L)\Big)^2} =\\
\sqrt{2\Big( \mathrm{Tr} (H^2 \rho) - \mathrm{Tr}(H\rho)^2  +\mathrm{Tr}(G^2\rho) -\mathrm{Tr}(G\rho)^2  + i \mathrm{Tr}([H,G]\rho) \Big )}.
\end{multline}
\end{linenomath}
For unitary evolution, $G=0$,  one obtains from~(\ref{eq:Speed0})
\begin{linenomath}
\begin{equation}\label{eq:SpeedA6}
\vartheta (L)  =
\sqrt{2\Big( \mathrm{Tr} (H^2 \rho) - \mathrm{Tr}(H\rho)^2\Big )}
\end{equation}
\end{linenomath}
while
\begin{linenomath}
\begin{equation}\label{eq:SpeedA7}
\vartheta_{\rm pure} (L)  =
\sqrt{2\Big( \mathrm{Tr} (H^2 \rho) - \mathrm{Tr}(H\rho)^2\Big) }
\end{equation}
\end{linenomath}
Finally, note that in all the above cases applied to the two-level quantum systems (two-flavour oscillation) described by a density matrix of the form~(\ref{eq:RhoNeutrinos})  an extremely simple formula  for $\vartheta$~(\ref{eq:Speed4}) results
\begin{linenomath}
\begin{equation}\label{eq:SpeedA8}
\vartheta_{\rm pure} (L)  =
\sqrt{ \tfrac{1}{2}  \bm{n}^{\,\prime}(L) ^2}.
\end{equation}
\end{linenomath}

\end{document}